\documentstyle[prl,eqsecnum,aps]{revtex}

\begin{document}

\title{ EFFECTIVE HAMILTONIAN IN THE PROBLEM OF A "CENTRAL SPIN" 
COUPLED TO A SPIN ENVIRONMENT }

\author{I. S. Tupitsyn$^{1}$, N. V. Prokof'ev$^{1,2}$, P. C. E. Stamp$^{2}$  }
\address{
$^{1}$ Russian Science Center "Kurchatov Institute", Moscow 123182, Russia\\
$\;\;\;$ \\
$^{2}$ Physics Department, University of British Columbia, 6224 Agricultural
Rd.,\\
 Vancouver B.C., Canada V6T 1Z1 \\ }
\maketitle

\vspace{1cm}
\begin{abstract}
We consider here the problem of a "giant spin", with spin quantum 
number $S \gg 1$, interacting with a set of microscopic spins. 
Interactions between the microscopic spins are ignored. This model 
describes the low-energy properties of 
magnetic grains or magnetic macromolecules (ferromagnetically or
antiferromagnetically ordered) interacting 
with a surrounding spin environment, such as nuclear spins. 

Our aim is to give a general method for truncating the model to
another one, valid at low energies, in which a two-level system
interacts with the environmental spins, and higher energy terms are
absorbed into a new set of couplings. This is done using an
instanton technique.

We then study the accuracy of this technique, by comparing the
results for the low energy effective Hamiltonian, with results
derived for the original giant spin , coupled to a macroscopic
spin, using exact diagonalisation techniques. We find that the low
energy central spin effective Hamiltonian gives very accurate
results (with increasing accuracy for large $S$), provided the
typical coupling energies between the giant spin and the
microscopic spins are not too large, and provided temperature and
external field are sufficiently low. The essential limitation to
the applicability of the low-energy effective Hamiltonian is just the 
semiclassical WKB approximation itself, which inevitably fails for
very small $S$.

Our results thus justify previous use of this effective Hamiltonian in 
calculations of the effects of nuclear spins on the dynamics of 
nanomagnetic systems.

\end{abstract}

\vspace{1cm}
\pacs{PACS numbers: 75.10.Jm, 03.65.Db, 75.60.Jp}

\section{Introduction}
\label{sec:in}

In this paper we examine a model which
was originally introduced to deal with the effects of 
environmental spins on the coherent tunneling of magnetic grain 
magnetisation \cite{Stamp92,our93,our-book}. It consists of a central 
"giant spin" \cite{Hemmen,Chudnovsky88,QTM94}, having
spin quantum number $S \gg 1$, coupled to a set $\{ {\vec \sigma}_k \}$ of 
"microscopic spins"; the microscopic spins are only very 
weakly coupled to each other. The model is then completely described 
by a Hamiltonian 
$H_o( {\vec S} )$ for the central spin, a Hamiltonian 
$H_{env} ( \{ {\vec \sigma}_k \} ) $ for the microscopic spins, and a 
coupling $H_{int} ( {\vec S} , \{ {\vec \sigma}_k \} )$ between the
two:
\begin{equation}
H(\vec{S} , \vec{\sigma}_k ) = H_o(\vec{S}) + {1 \over 2S}
\sum_{k=1}^N \omega_k \vec{S} \cdot \vec{\sigma}_k +
V_{dip} (\{ \vec{\sigma}_k ,\vec{\sigma}_{k^\prime } \} ) \;,
\label{a.1}
\end{equation}
in which we specialise immediately to the form relevant to 
most of this paper, in which the interaction between the central
spin and the environment has a hyperfine form, and the dynamics of the 
environmental spins (considered now to be nuclear spins), comes from their
mutual dipolar or indirect Suhl-Nakamura interactions.
In most practical applications of this "central spin model", the environmental
nuclear spins will be either inside the object
carrying the giant spin, or near it, in some substrate, or solvent, or 
surrounding matrix.
The central spin may be 
a magnetic grain \cite{Chudnovsky88} or a 
magnetic macromolecule such as ferritin 
\cite{Awschalom}
or, on a smaller scale \cite{Villain}, $ Mn_{12}$-ac. 
Similar models were also introduced some time ago to describe the 
large superparamagnetic "spin clusters" which are believed 
to exist in many disordered magnets at low temperature, such as 
$Si:P$ near the metal-insulator transition \cite{Sachdev}, or "giant magnetic 
polarons" \cite{Kasuya}. Similar spin clusters exist in "quantum spin glasses"
\cite{Sachdev94}. In these models the giant spin is somewhat less uniquely
defined (indeed its size depends on temperature). Finally,
a similar central spin model was considered a long time ago by
Gaudin, who wrote down a set of Bethe ansatz equations for it
\cite{Gaudin}.

We should also note that models like (\ref{a.1}) can be
of interest even for non-magnetic systems (although the coupling
term will not then be a hyperfine one). This is 
because most large quantum objects will have (at least) a coupling to 
their own nuclei, and perhaps to other quasi-free spin degrees of freedom
like paramagnetic impurities. 
Thus in the oft-cited example of the superconducting 
SQUID ring \cite{Caldeira}, there is an electromagnetic coupling of the 
tunneling flux coordinate (and the associated ring supercurrent) to all 
of the nuclear spins within a penetration depth of the surface (either 
in the junction or in the ring), as well as to any paramagnetic impurity 
spins that happen to be present \cite{Stamp88}. The nuclear spins
can then not only change the energetics of the relevant collective
coordinate, but also dephase it,
in a kind of ``inverse Stern-Gerlach effect", where nuclear spins
play the role of measuring devices \cite{Stamp88,Stamp-review}. It then 
becomes important to determine how the coupling to the background 
spins will affect the quantum dynamics of the central system. This
depends of course on the system; for example, in the case of
SQUID's, it is fairly easy to show that the effect of the
environmental nuclear and paramagnetic spins on flux tunneling is
usually negligible, but that the effect on ``Macroscopic Quantum
Coherence" may not be. On the other hand, the strength of the
hyperfine coupling, between the order parameter of many
nanomagnetic systems and their nuclear spins, is such that the
nuclear environment can seriously disrupt the quantum dynamics of
the nanomagnet at low $T$.

The purpose of the present paper is not to analyze the effects of
the environmental spin bath on the dynamics of nanomagnet - we have
done this in a number of other papers. Instead we wish to
investigate the validity of the effective Hamiltonian we have
previously employed. This Hamiltonian results from a truncation to
a  lower-dimensional Hilbert space in which the giant spin dynamics
are reduced to that of a two-level system (still coupled to $N$
environmental spins). This truncation is analogous to the
truncation of a quantum system like a SQUID, coupled to delocalized
electrons, to the
``spin-boson" model at low $T$. In the present case of the central
spin, the truncated central spin effective Hamiltonian takes the
form
\begin{eqnarray}
H_{eff} & = & \left\{ 2 \Delta_o \hat{\tau }_ -  \cos \bigg[ \pi S +
 \sum_k \alpha_k \vec{n} \cdot \hat{\vec{\sigma }}_k  -
 \beta_o \vec{n} \cdot \vec{H}_o  \bigg] + H.c. \right\} \nonumber \\
& + & {{\hat \tau }_z \over 2} \sum_{k=1}^N \omega_k^{\parallel} \:
 {\vec l}_k \cdot {\hat {\vec \sigma }}_k  + {1 \over 2} \sum_{k=1}^N 
\omega_k^{\perp}\: {\vec m}_k \cdot {\hat {\vec \sigma }}_k 
+\sum_{k=1}^N \sum_{k'=1}^N V_{kk'}^{\alpha \beta } \hat{\sigma}_k^\alpha 
\hat{\sigma}_{k'}^\beta ~ \;.
\label{z.301}
\end{eqnarray}
In this Hamiltonian the two-level dynamics of $S$ will be described
by the Pauli vector matrix $\hat{\vec{\tau}}$, and we have assumed
that $H_o(\vec{S})$ in (\ref{a.1}) contains an external field
$\vec{H}_o$. This is the same form as we have used in previous
investigations \cite{our-book,JLTP}. Here $\Delta_o$ is a tunneling
matrix element, $\vec{n}$, $\vec{l}_k$, and $\vec{m}_k$ are unit
vectors, with $\vec{l}_k$ and $\vec{m}_k$ mutually perpendicular;
$\alpha_k$ and $\beta_o$ are dimensionless parameters,  and
$\omega_k^\parallel$ and $\omega_k^\perp$ are ``longitudinal" and
``transverse" diagonal coupling energies \cite{JLTP}.

In the present paper we will show how one goes from (\ref{a.1}) to 
(\ref{z.301}), in an instanton calculation; and we will discuss how
accurately (\ref{z.301}) represents the dynamics of (\ref{a.1}),
by carrying out exact diagonalization studies of (\ref{a.1}), and
comparing them with the predictions of (\ref{z.301}). To do this we
will pick a special (but actually quite generic) form for
$H_o(\vec{S})$, viz., a biaxial (easy axis/easy plane) form 
\begin{equation}
H_o( {\vec S} ) = {1 \over S} \bigg[ -K_{\parallel}\:S_z^2 
+ K_{\perp}S_y^2 \bigg] \;, 
\label{1}
\end{equation}
and calculate the parameters in (\ref{z.301}) deriving
from ({\ref{1}), in an instanton expansion. We 
should note here that some similar
exact diagonalisation studies of this model 
were recently carried out by
Abarenkova \& Angles d'Auriac \cite{d'Auriac}, who also 
compared their exact diagonalisation results with a somewhat simplified 
version of (\ref{z.301}) in certain limits. In the present work we have 
a rather different aim; we wish to show that the structure of the 
exact diagonalisation results can be put in a form very close to that in 
(\ref{z.301}), and then we can check numerically whether the coefficients
in (\ref{z.301}) agree with those we calculate using an instanton procedure,
over a very wide paramenter range. Thus we are not just trying to test the
truncated model numerically; we are also trying to test its {\it form}, and
to see if anything is missing from it. 

We find that, as assumed in previous work, {\ref{z.301}) gives a
very good description of the low-energy dynamics of (\ref{a.1}),
both as regards the mathematical form and the numerical values calculated
for the coefficients in an instanton expansion,
{\it provided} that 

(i) The semiclassical approximation for the ``bare" giant spin
Hamiltonian $H_o(\vec{S})$ is a good one;

(ii) The parameter $E_o=\omega_o N^{1/2}$, characterizing the
spread in energies caused by the hyperfine coupling, and the
thermal energy $k_BT$, satisfy the inequality $E_o, k_BT \ll
\Omega_o$, where $\Omega_o$ is the ``bounce frequency" associated
with semiclassical tunneling, which characterizes roughly the
energy gap between the lowest doublet of states of $H_o(\vec{S})$,
and the higher excited states. Thus the result (\ref{z.301}) can
start to fail if either the mean hyperfine coupling $\omega_o$, or
the number $N$ of nuclear spins, becomes too large.

(iii) The ``bias" energy $\xi_H =\gamma_e
(\vec{S}_1-\vec{S}_2)\cdot \vec{H}_o$, where $\vec{S}_1$ and
$\vec{S}_2$ are the two semiclassical minima of $H_o(\vec{S})$,
must also satisfy $\xi_H \ll \Omega_o$.

Conditions (ii) and (iii) simply prevent the two lowest
levels of $H_o(\vec{S})$ from significantly mixing with higher
levels, once the nuclear spins or magnetic field are coupled in.

Although these results are derived for the particular example
(\ref{1}), we believe that the instanton method we use can be
applied to most reasonable forms of $H(\vec{S}, \{ \vec{\sigma} \}
)$ like (\ref{a.1}), to give a truncated form like (\ref{z.301}),
provided
the same ``low-energy" restrictions (i)-(iii) are satisfied; all
that will change  will be the values of the parameters in
(\ref{z.301}) (for the case (\ref{a.1}), the 
values of the parameters appearing in (\ref{z.301}) are given in
Sec. III below).

In what follows we begin (in section \ref{sec:2}) by describing the 
central spin model, and giving some discussion of 
the relevant physical coupling energies in it. 
In section \ref{sec:3} we show how an initial microscopic Hamiltonian,
describing the giant central spin and its microscopic environmental
satellites, can be truncated to a low-energy effective Hamiltonian description.
In section \ref{sec:4} we proceed to the discussion of the results of
the exact diagonalization studies and, finally, in section \ref{sec:5}, 
we summarize our results.

\section{The model, and some simple examples}
\label{sec:2}

In this section we briefly describe the physical origin 
of the central spin model
for nanomagnetic systems, and the Hamiltonian of eqtn. (\ref{a.1}),
and also remark on what is left out of it. Since the variety of
physical
examples is rather large, as is the range of possible coupling
energies, we also mention some reference examples.

Let us consider first the central spin itself, whose dynamics is
described in eqtn. (\ref{a.1}) by $H_o({\vec S})$.
A typical central spin is made up of a very large number of microscopic spins,
whose motion is assumed to be
locked together by nearest-neighbour exchange, or ``Hund's rule",
 couplings (or, in the case 
of superparamagnetic spin complexes, by indirect spin-spin 
couplings such as the RKKY interaction). These couplings can be very
large, as much as $0.1-1\: eV$; this is much larger than other individual spin
energy scales such as those coming from anisotropy, so that in principle
one can have large monodomain magnetic particles, with many spins (up to 
perhaps $10^8$, depending on the particular systems) lined up to form
a "giant spin"\cite{Hemmen,Chudnovsky88,Stamp-review}.
 Another more subtle example is the 
antiferromagnetic grain, in which  nearest neighbours are anti-parallel, 
and one obtains a "giant N\'{e}el vector" \cite{Barbara90,Zaslavsky}. 
The superparamagnetic 
spin complexes are more delicate still, and typically arise at low 
temperatures when paramagnetic electronic spins in a disordered magnet 
"lock together" via indirect exchange, RKKY interactions, 
etc \cite{Sachdev}.

The first step in analysing such a system is to treat the entire
spin complex as a rigid quantum rotator ${\vec S}$, with 
dynamics governed by an  anisotropy field in the 
particle/grain; i.e., write a central spin "bare Hamiltonian"
$H_o( {\vec S} )$, with $\mid {\vec S} \mid = S$ a constant. 
The next step to take, if the temperature is low, is to try truncating 
a Hamiltonian like (\ref{1}) to a 2-level system, involving only the lowest 
2 levels. This has been done by various authors;
in fact the papers of van Hemmen and Suto \cite{Hemmen} analyzed a much more 
general class of bare central spin Hamiltonians than (\ref{1}).

The giant spin model omits the  higher excited states of the particle, 
involving spin flips of spins inside the grain (internal magnons); 
the neglect of these is usually justified  by appeal to the large 
value $\mid J \mid$ of spin exchange \cite{Chudnovsky88,Stamp-review}, 
in models like 
equation (\ref{1}). Incorporation of such processes implies 
relaxation of the constraint  $\mid {\vec S} \mid$=constant.
In most circumstances this will hardly be a serious omission; the
lowest ``internal magnon" excitations do not even reach the top of
the ``ginat spin" multiplet of $(2s+1)$ levels until the grain or
molecule reaches a radius $L \sim (Ra)^{1/2}$, for an AFM grain ($R$ is the
domain wall length, roughly equal to the maximum size of a
monodomain magnet; and $a$ is the typical interspin spacing in the
magnet); or $L \sim (R^2a^3)^{1/5}$ for a FM grain. 
Even when we do get mixing of internal magnon modes and giant spin
levels at these energies (i.e., energies $\sim 2SK$, where $K$ is
a typical anisotropy term), this will hardly affect either the
energetics or dynamics at energy scales $\Omega_o \sim K$ or below.
Another 
omission is that of surface excitations of the system - in any real 
nanomagnetic grain, there can be low-energy magnetic excitations involving 
"loose spins", i.e., spins which, because of imperfect preparation 
or the inevitable defects, dislocations, etc., have a coupling 
energy $\ll J$ to their neighbours. These couplings are hard to
quantify, but it is rather unlikely that they will be as small as 
the single-ion anisotropy energies in (\ref{1}). We will also
ignore the possibility that surface spins might have anisotropy 
energies differing from those inside the grain.

Finally, in this paper we will ignore all couplings to external
bosonic baths such as electrons, phonon, or photons. Electron
effects on nanomagnets were recently discussed by us \cite{JLTP}
(as well as by Fukuyama and Tatara for domain walls \cite{Tatara}).
Phonon effects have been discussed by many authors 
\cite{GargKim,Chudnovsky94}. Photon effects were shown to be
negligible some time ago \cite{Stamp91}. Experimental predictions
for the behavior of nanomagnets coupled simultaneously to
oscillator and spin bath were given in Ref. \cite{JLTP}; the work
was also recently
applied to the Mn$_{12}$-acetate system \cite{mn12}.

We now turn to the problem of main interest here, i.e.,  
the coupling of ${\vec S}$ to other spin degrees of freedom - as already 
noted, we are primarily interested in nuclear spins, 
both inside and outside the grain. 
A nucleus with a finite spin $I$, and spin moment
$-\gamma_N \hbar I$, interacts both locally with the 
electron clouds at the same ionic site, via the contact hyperfine
interaction, and also non-locally with other ions via the dipolar field.
There may also be other residual interactions such as transfer hyperfine 
interaction (as for, e.g., $F^{19}$ in $MnF_2$ or $CoF_2$), or quadrupolar 
coupling if $I>1/2$. Finally, one may have interactions between the different
nuclear spins, via, e.g., spin waves (the Suhl-Nakamura interaction),
or dipolar couplings - these are very weak 
(Suhl-Nakamura interactions are $\sim 10^{-6} \: K$, and nuclear dipole-dipole 
interactions $\sim 10^{-7} \: K$) but, it turns out, can play a role in
the dynamics of $\vec{S}$. 

The most important interactions are the contact hyperfine interaction 
$ -\gamma_N \hbar {\vec I} \cdot  {\cal A} \cdot {\vec S}_o $ to
the electronic spin on the same ionic site,
and the dipolar interaction between the nuclei and electrons.
The strength of these interactions is measured directly in NMR experiments;
the hyperfine couplings range from $1-3\:MHz$ (i.e., $5-15\times 10^{-5}K$)
for protons in $H^1$, up to values greater than $5000\:MHz$ ($0.25\:K$) for
some rare-earth magnetic nuclei (e.g., $Tb^{159}$, $Dy^{163}$); these
latter correspond to local fields acting on the nuclei as high as $500 \:
Tesla$, and come overwhelmingly from the contact interaction. Thus when
a central spin rotates, with all internal electronic spins rotating in unison,
the contact interaction tries to force the nuclei on the magnetic sites to 
follow. If the contact hyperfine interaction strength is $\omega_o$, then it is 
clear that the ratio $\omega_o /\Omega_o$, with $\Omega_o^{-1}$ the 
timescale of central spin rotation, is going to be crucial to the 
nuclear spin dynamics. The parameter $\Omega_o$ for the
ferromagnetic grain is defined by the anisotropy energy 
$\Omega \sim K$, while for the antiferromagnetic grain it is much 
higher, typically $\Omega \sim 2\sqrt{KJ} $
\cite{Barbara90,Zaslavsky}.
 If the electronic spin frequencies $\Omega_o \gg 
\omega_o$, then the nuclear spins will experience a "sudden" perturbation, 
and few of them will follow the central spin; conversely, if $\omega_o \gg 
\Omega_o$, the nuclear spins will follow adiabatically. In fact, the ratio 
$\omega_o /\Omega_o$ is usually somewhere between $0.001-0.1$. 
On the other hand for nuclei in non-magnetic
ions, $\omega_o /\Omega_o$ may be much smaller, and will of 
course depend strongly on the host (in a magnetic host, 
transfer hyperfine interactions 
$ -\gamma_N \hbar {\vec I} \cdot  {\cal A}_{0j} \cdot {\vec S}_j $
from nearby magnetic moments ${\vec S}_j$, can greatly increase $\omega_o$;
for example, $\omega_o$ for $F^{19}$ in $MnF_2$ is $160\: MHz$).

A large central spin will also interact with nuclear spins in 
the surrounding  medium (a substrate or solvent) via the dipolar 
interaction generated by the central spin dipolar field.
This interaction may not be 
negligible; for a central spin with $S=10^4$, the dipolar coupling to a 
nucleus
at a distance of $100 \AA$ is already $\sim 1\:MHz$, rising to
$30\: MHz$ at a distance of $30 \AA$; and there is clearly a very large 
number of nuclear spins within $100 \AA$ of the central spin! We
should  also note, that the same considerations may be also applied
to quite distant paramagnetic impurities. For $S=10^4$ they couple to
the giant spin in the $MHz$ range up to distances $1000 \AA$.

The approach we shall take is to include first the overwhelmingly dominant 
hyperfine interaction in the coupling between $S$ and each nuclear spin.
Since the individual electronic moments are locked together this leads
directly to the second term in (\ref{a.1}). We will drop the
internuclear couplings; these are so weak that their effects can be
handled in a very simple way once the main ``central spin" problem
has been solved, as a weak after-effect \cite{JLTP}. 
In fact it is trivial to show that the effects of the 
internuclear dipolar coupling can be entirely handled, in this
effective Hamiltonian approach, by adding back the dipolar coupling
to the low-energy effective Hamiltonian in exactly the same form as it 
appears in the high-energy Hamiltonian- this is what we have done in
writing equations (\ref{a.1}) and (\ref{z.301}). Thus, we shall use,
as our starting point for analysis, the model in Eq.~(\ref{a.1}), without the
internuclear dipolar interaction.

\section{The effective Hamiltonian}
\label{sec:3}

The task of this section is to truncate the "giant spin" description,
 and find a new effective  Hamiltonian which operates
only at low energies. In fact we reduce the $(2S+1)$-dimensional Hilbert space
of the giant spin itself down to two lowest levels - 
the giant spin is now described
by a Pauli vector matrix $\vec{\tau }$. This two-level system interacts with
a "spin-bath", described by Pauli matrices $\{ \hat{\vec{\sigma}}_k \}$, with
$k=1,2,3, \dots N$. The new effective Hamiltonian thus operates in a reduced
Hilbert space, of complex dimension $2^{N+1}$. We may use this Hamiltonian
$H_{eff}$ {\it provided} higher energy degrees of freedom are not excited.
This means in practice that $H_{eff}$ is restricted to energy scales 
$\ll \Omega_o$, the characteristic frequency of zero point
fluctuations of $\vec{S}$. The higher levels do
of course affect $H_{eff}$; in the usual way, their effects are incorporated
into renormalised "coupling constants" in $H_{eff}$.

In what follows we first explain how the truncation procedure is done for
this kind of problem; then we derive $H_{eff}$, starting from the giant spin model.

\subsection{ Free Giant Spin}

The technique we use is the standard one of separating the "slow" and "fast"
degrees of freedom in the problem; for our giant spin the difference between
the two is illustrated in Fig. 1. The truncated Hamiltonian should only
include slow variables, the fast ones having been eliminated in favour of a
new set of couplings between the slow ones.

Let us first briefly recall how the truncation is done in the 
{\it absence} of the environmental spins and magnetic field
- this makes the method clear, and also gives
us some results we will need later. Since $S \gg 1$, one uses semiclassical
methods - the task is to start from a free giant spin Hamiltonian 
$H_o(\vec{S})$ like (\ref{1}), and derive a reduced Hamiltonian 
$H^o_{eff} (\vec{\tau})$, operating in the space of the lowest two 
levels of $H_o(\vec{S})$. We shall choose a basis in this truncated 
space such that the eigenstates of $\hat{\tau}_z$ correspond to the 
2 semiclassical minimum states of $H_o(\vec{S})$, defined by coherent
state vectors $\vert \vec{n}_1 \rangle $ and $\vert \vec{n}_2 \rangle $,
such that 
$\langle \vec{n}_1 \vert \vec{S} \vert \vec{n}_1 \rangle = S \vec{n}_1$
and
$\langle \vec{n}_2 \vert \vec{S} \vert \vec{n}_2 \rangle = S \vec{n}_2$;
the eigenstates of $H_{eff}^o(\vec{\tau})$ are then linear combinations of 
$\vert \vec{n}_1 \rangle $ and $\vert \vec{n}_2 \rangle $,
which we can determine once we have found the four matrix elements
$\langle \vec{n}_{\alpha} \vert H^o_{eff}
\vert \vec{n}_{\beta } \rangle $ with $\alpha ,\beta =1,2 $.

Formally one can do this as follows, for the free spin. Consider the path
integral expression for the transition amplitude 
$\Gamma^o_{\alpha \beta } (t)$,
during the time $t$; this is given by \cite{Chudnovsky88,Stamp-review}:
\begin{equation}
\Gamma^o_{ \alpha \beta }(t) = \langle \vec{n}_{\alpha} \vert
e^{-iH_o(\vec{S})t} \vert \vec{n}_{\beta } \rangle = 
\int_{\vec{n}(\tau =0 )= \vec{n}_{\beta } }^{\vec{n} (\tau=t) 
=\vec{n}_{\alpha}} 
{\cal D} \vec{n} (\tau ) \exp 
\left\{ -\int_0^t d \tau {\cal L}_o (\tau ) \right\} \;,
\label{z.1}
\end{equation}
where the free spin  Euclidean Lagrangian is 
\begin{equation}
{\cal L}_o = - iS \dot{\theta } \varphi \sin \theta + H_o(\vec{n} )\;.
\label{z.2}
\end{equation}
Here $\theta$ and $\varphi$ are the usual polar and 
azimuthal angles for the unit vector  $\vec{n}(\tau )$. 

Now in the semiclassical approximation there are two fundamental time scales
in the paths $\vec{n} (\tau )$ in (\ref{z.1}); these are $\Omega_o^{-1}$, the
time required for the instanton traversal to be made between states, and
$\Delta_o^{-1}$, the typical time elapsing between instantons. 
By definition, an effective Hamiltonian is supposed to reproduce
the slow dynamics of the system in the truncated Hilbert space of
the two lowest levels, i.e., for long time scales an evolution
operator is approximated as
\begin{equation}
\Gamma^o_{ \alpha \beta }(t) \approx \left( e^{-iH_{eff} t}
\right)_{ \alpha \beta }   \;.
\label{evolution}
\end{equation}
Since $\Delta_o$ is exponentially smaller than $\Omega_o$, and
nondiagonal elements are $H_{eff} \sim \Delta_o $, we can write
\begin{eqnarray}
\Gamma^o_{\alpha \beta  }(t) &= &\langle \vec{n}_{\alpha} \vert
e^{-iH_{eff} t} \vert \vec{n}_{\beta } \rangle \nonumber \\
 & \approx &  \delta _{\alpha \beta } -it 
\langle \vec{n}_{\alpha} \vert H^o_{eff} \vert \vec{n}_{\beta } \rangle \;;
\;\;\;\;
(\Omega_o^{-1} \ll t \ll \Delta_o^{-1}) \;;
\label{z.3}
\end{eqnarray}
Then we immediately find the matrix elements of $H^o_{eff} (\vec{\tau })$
for $\alpha \ne \beta $  as
\begin{equation}
\left( H^o_{eff}(\vec{\tau }) \right)_{\alpha \beta  } = {i \over t} 
\Gamma^o_{\alpha \beta  }(t)\;; \;\;\;\;
(\Omega_o^{-1} \ll t \ll \Delta_o^{-1})\;.
\label{z.4}
\end{equation}

As a concrete example, consider the easy-axis/easy-plane Hamiltonian
(\ref{1}), where
\begin{equation}
H_o(\vec{n})= S K_{\parallel} \bigg[ \sin ^2\theta + 
{ K_{\perp} \over K_{\parallel} }\: \sin ^2\theta  \sin ^2 \varphi \bigg] \;,
\label{z.5}
\end{equation}
The two lowest states  are $\vec{n}_{1}=\hat{\vec{z}}$
and  $\vec{n}_{2 }=-\hat{\vec{z}}$; henceforth we write these states 
as $\vert \Uparrow \rangle $ and  $\vert \Downarrow \rangle $.
 In the usual case where 
$K_{\perp} / K_{\parallel} \gg 1$ (so that the tunneling amplitude is 
appreciable) one has only small oscillations of $\varphi $ about the 
semiclassical trajectories $\varphi =0$ or $\pi$, and by eliminating $\varphi $
one has 
\begin{equation}
{\cal L}_o (\theta ) = { S \over 4 K_{\perp} } \dot{\theta }^2 +
S K_{\parallel} \sin ^2\theta \;,
\label{z.6}
\end{equation}
giving a classical equation of motion $\dot{\theta } = \Omega_o \sin \theta $,
and instanton solution \cite{Chudnovsky88,Stamp-review}, 
going from $\vert \Uparrow \rangle $ to $\vert \Downarrow \rangle $
\begin{equation} 
\sin \theta (\tau ) =1/ \cosh (\Omega_o \tau )
\label{z.sin}
\end{equation}
(centered at $\tau =0$);
in this system
\begin{equation}
\Omega_o = 2 (K_{\parallel}K_{\perp})^{1/2}  \;. 
\label{z.7}
\end{equation}

The frequency $\Omega_o$ then sets the ultraviolet cut-off for the Hilbert
space of $H^o_{eff} (\vec{\tau })$, and one finds, by substituting the 
semiclassical solution into (\ref{z.1}) and evaluating a determinant over 
the quadratic fluctuations around the semiclassical solution
\cite{Callan} (the zero mode contribution in the usual way gives a
factor $it$), that from (\ref{z.4}) we get
\cite{Chudnovsky88,Stamp-review,Hemmen,Ioselevich,Delft}: 
\begin{equation} 
\hat{H}^o_{eff} (\vec{\tau }) = \Delta_o ( S ) \hat{\tau}_{x}  \;,
\label{z.8}
\end{equation}
\begin{equation}
\Delta_o ( S ) = -\sum_{\eta =\pm }\sqrt{{2 \over \pi} Re A_o^{(\eta )}}
\Omega_o \exp \{ -A_o^{(\eta )} \} \equiv 2\Delta_o \cos \pi S \;,
\label{z.9}
\end{equation}
\begin{equation} 
A_o^{(\eta )} = 2S ( K_{\parallel}/K_{\perp} )^{1/2} + i\eta \pi S  \;,
\label{z.10}
\end{equation}
where the action $A_o^{(\eta )}$ is that for transition between the two
semiclassical minima, either clockwise ( $\eta =+$) or anticlockwise 
($\eta =-$); the phase $\eta \pi S $ is the Kramers/Haldane phase,
coming from the linear in time derivatives kinetic term in (\ref{z.2}).
Without this phase, we would simply have a splitting 
$\vert \Delta_o \vert  = \sqrt{2  Re A_o/\pi } \Omega_o \exp \{ -A_o \}$ 
with $ A_o = 2S ( K_{\parallel}/K_{\perp} )^{1/2}$.


\subsection{ Giant Spin in External Magnetic Field}

Let us now include a weak magnetic field 
($h = \gamma_e \mid {\vec H}_o \mid /2K_{\parallel} \ll 1$). 
The effect of an applied ${\vec H}_o$ 
is mostly nicely seen in the instanton language; since the 2 possible 
paths between the degenerate minima involve opposite Haldane topological 
phase \cite{Haldane,Delft}, this phase can be changed by an external field 
\cite{GargEL}, causing the tunneling splitting to change and even oscillate.
Then the 
corresponding Lagrangian written in terms of angles $(\theta, \varphi)$
has the form (in imaginary time)
\begin{eqnarray}
{\cal L} & =& {\cal L}_o + \delta {\cal L}_h   \nonumber \\
\delta {\cal L}_h & =& -\gamma_e S ( H_o^x \sin \theta \cos \varphi + 
H_o^y \sin \theta \sin \varphi + H_o^z \cos \theta ) \;,
\label{h.99}
\end{eqnarray}
where ${\cal L}_o$ was defined in (\ref{z.2}). As before, we perform
the Gaussian integration over the small deviations of $\varphi$ around zero
or $\pi$, and obtain the effective  Lagrangian for $\theta$. Keeping
only linear terms in magnetic field we  write 
\begin{equation}
\delta {\cal L}_h (\theta ) = -\gamma_e S ( H_o^x \sin \theta + 
{i \dot{\theta }\over 2K_{\perp }} H_o^y + H_o^z \cos \theta ) \;,
\label{h.98}
\end{equation}
whereas ${\cal L}_o (\theta)$ has the form (\ref{z.6}). 
With the same accuracy in $h \ll 1$, the correction to the effective 
action is given by 
\begin{equation} 
A^{(\eta )} =A_o^{(\eta )} + \int d\tau \delta {\cal L}_h
 ({\vec S}_o^{(\eta )} (\tau ))\;,
\label{h.97}
\end{equation}
where ${\vec S}_o^{(\eta )} (\tau )$ is the semi-classical trajectory 
defined by (\ref{z.sin}). Substituting (\ref{z.sin}) in(\ref{h.97}) we find 
\begin{equation}
\delta A_h^{(\eta)} = -\eta { \gamma_e \pi S \over \Omega_o } 
(\hat{\vec{x}} + i \sqrt{K_{\parallel} / K_{\perp }} \hat{\vec{y}} ) 
\cdot {\vec H}_o  \;,
\label{h.96}
\end{equation}
and the final expression for the effective action in  the weak 
magnetic field takes the form \cite{GargEL}
\begin{equation}
A^{(\eta )} = 2S \sqrt{ K_{\parallel} \over K_{\perp} } -  
\eta { \gamma_e \pi S \over \Omega_o } H_o^x + 
i \eta \left( \pi S - { \gamma_e \pi S \over \Omega_o }
\sqrt{K_{\parallel} \over K_{\perp }} H_o^y \right)  \;,
\label{h.95}
\end{equation}
If ${\vec H}_o$ is directed along ${\hat {\vec z}}$ ({\it easy-axis}),
this {\it biases} the symmetric 2-well problem in (\ref{1}). Within 
our accuracy (we kept only linear terms in ${\vec H}_o$) the correction 
to the topological phase due to $H_o^z$ iz zero. However if applied along 
${\hat {\vec x}}$, the magnetic field lowers the barrier height and displaces 
the degenerate minima towards each other in the XZ-plane. In this case 
the clockwise and anticlockwise paths become inequivalent and the 
tunneling splitting changes with the magnetic field. Application 
of ${\vec H}_o$ along ${\hat {\vec y}}$ lowers the barrier and distorts 
the semi-classical path. This field leads to oscillations in the value 
of the tunneling splitting \cite{GargEL} since  the second 
term in the parentheses is nothing but the Berry phase accumulated by 
the giant spin during the tunneling.


\subsection{ Giant Spin plus Spin Bath}

We now include the interaction of the giant spin with the 
environmental spins. As discussed  in section II, we start with the
 Hamiltonian (\ref{a.1}) (after dropping $V_{dip}$),
  where the $\{ \omega_k \}$ 
are the hyperfine couplings at each site, and the $N$ Pauli variables 
$\{ \vec{\sigma}_k \}$ describe each of the spin-1/2 bath modes. 
Without the hyperfine term, the total nuclear spectrum, containing
$(2)^N$ lines, is almost completely degenerate, with only a tiny spreading  
$\sim T_2^{-1}$ of levels caused by the internuclear dipolar interaction 
$V_{dip}$. With the hyperfine interaction, the nuclear levels are now spread
over a total range $\sim \omega_o N$ (assuming $\omega_k \sim \omega_o $ 
for all nuclei), with most levels concentrated in a Gaussian peak of 
half-width $\sim \omega_o N^{1/2} $. Since $\omega_o \gg T_2^{-1}$, 
this means that the nuclear spin spectrum and dynamics are basically
{\it slaved } to the spectrum and dynamics of $\vec{S}$. This of course 
sharply contrasts with the usual oscillator bath environments, where the 
bath spectrum is only  very weakly perturbed by the coupling to a macroscopic 
system.

What we wish to do is to truncate to a low energy effective Hamiltonian
$H^o_{eff} (\vec{\tau }, \{ \vec{\sigma}_k \})$, where $\vec{\tau }$, 
as before, operates in the subspace of the lowest two levels of $\vec{S}$.
The general way to do this is to separate "slow" (frequency $\ll \Omega_o$)
from fast (frequency $\gg \Omega_o$) processes, in the combined 
system/environment dynamics. One then incorporates the fast processes 
into parameters of the low-energy effective Hamiltonian. In the instanton 
method, there are two kinds of processes that one may consider, shown 
schematically in Fig. 2. In the periods between instantons (i.e., 
when $\vec{n} = \vec{n}_1$ or $\vec{n}_2$), there are {\it diagonal 
couplings} between the environment and the system - in general,
if one takes into account the bath dynamics, these will be dynamical. On 
the other hand, the transition between $\vec{n}_1$ and  $\vec{n}_2$ (the 
instanton) can obviously perturb the environment - thus there must, in 
the reduced Hamiltonian, be a dynamic {\it non-diagonal} coupling 
generated between the system and environment. The large values 
of $\omega_k$ mean that {\it during} the instanton transitions of $\vec{S}$, 
a great deal can happen in the nuclear bath - the time-varying hyperfine 
field $\omega_k \vec{S}(\tau ) /S$, acting on each $\vec{\sigma}_k$ during 
the transition, can cause $\vec{\sigma}_k$ to flip. Since one typically 
has $\omega_k /\Omega_o \ll 1$ (weak coupling regime), the probability 
that a particular $\vec{\sigma}_k$ will flip during a single instanton 
passage is $ \vert \alpha_k \vert ^2/2 $, where $\vert \alpha_k
\vert \sim \pi \omega_k /2\Omega_o$
(this result is derived below, in the course of our derivation of $H_{eff}$;
see also refs. \cite{Stamp92,our-book}). 
Thus in each tunneling transition, roughly 
$\lambda = 1/2 \sum_k \vert \alpha_k \vert ^2 $ nuclear spins flip, and often 
$\lambda >1$.
Thus in the coupling to the spin bath we must deal
with {\it dynamic} interactions between $\vec{S}$ and the $\vec{\sigma}_k$,
during {\it the instanton itself}, when making the truncation to $H_{eff}$.
These will yield terms non-diagonal in the giant spin basis, in which $\tau_+$
or $\tau_-$ may be coupled to several environmental spins at once.

{\bf Diagonal coupling}: $\;\;$ 
We first deal with the truncated {\it static } interaction which must exist
between $\vec{\tau }$ and the $\{ \vec{\sigma}_k \}$ when no tunneling is
taking place; this is of necessity a diagonal term. Let the two
 relevant orientations of $\vec{S}$ are $\vec{S}_1$ 
and $\vec{S}_2$, with $\vec{S}_1 = -\vec{S}_2$. Then the only interaction 
with the spin bath is the hyperfine interaction, and the static contribution 
to the effective Hamiltonian is 
\begin{equation}
H_{eff}^D = 
{1 \over 2 } \hat{\tau}_z \sum_k \omega_k \hat{\sigma}_k^z \;, 
\label{3.94a}
\end{equation}
in the reduced Hilbert space.

In the more general case where $\vec{S}_1$ and $\vec{S}_2$ are not
antiparallel, we can still define a basis where $\vert \vec{n}_1 \rangle =
\vert \Uparrow \rangle $ and $\vert \vec{n}_2 \rangle =
\vert \Downarrow \rangle $. Now for the two different
orientations $\vec{S}_1$ and $\vec{S}_2$, the effective fields acting on
$\vec{\sigma}_k$ will be $\vec{\gamma}_k^{(1)}$ and  $\vec{\gamma}_k^{(2)}$.
These fields are the sums of the hyperfine fields $\omega_k \vec{n}_{1,2} $
and any other static fields acting on the nuclei. We  define the {\it sum} 
and the {\it difference} terms as
\begin{eqnarray}
\omega_k^{\parallel}{\vec l}_k & =& {\vec \gamma }_k^{(1)} - 
 {\vec \gamma }_k^{(2)} \nonumber \\
\omega_k^{\perp} {\vec m}_k & =& {\vec \gamma }_k^{(1)} + 
 {\vec \gamma }_k^{(2)}\;.
\label{3.88}
\end{eqnarray}
where the ${\vec l}_k$ and ${\vec m}_k$ are unit vectors. Then the truncated
diagonal interaction  takes the form (we project on states 
$\vert \Uparrow \rangle $ and $\vert \Downarrow \rangle $ using
standard $(1+\hat{\tau}_z)/2$ and $(1-\hat{\tau}_z)/2$ operators)
\begin{equation}
 H^D_{eff} = \sum_{k=1}^N \bigg\{ {\vec \gamma}_k^{(1)}
 {1+\hat{\tau}_z \over 2} + {\vec \gamma}_k^{(2)}
 {1-\hat{\tau}_z \over 2} \bigg\} \cdot {\hat {\vec \sigma }}_k
 \equiv   {1 \over 2} \bigg\{  
{\hat \tau }_z \sum_{k=1}^N \omega_k^{\parallel} \:
 {\vec l}_k \cdot {\hat {\vec \sigma }}_k  + \sum_{k=1}^N 
\omega_k^{\perp}\: {\vec m}_k \cdot {\hat {\vec \sigma }}_k \bigg\} \;,
\label{3.94}
\end{equation}
i.e., one term which changes when $\vec{S}_1 \to \vec{S}_2$, and one 
which does not. Usually $\omega_k^{\parallel} \gg \omega_k^{\perp}$ , and 
$\omega_k^{\parallel} \sim \omega_o$.

{\bf Non-diagonal coupling}: $\;\;$ 
Let us now turn to the dynamic interactions which occur when $\vec{S}$ 
is tunneling. There are two effects we must deal with - the effect of the 
nuclear spins on the giant spin dynamics during  tunneling, and also the 
effect of the motion of $\vec{S}$ on the nuclear spins themselves. In 
general both effects have to be handled together, in a mutually consistent 
way.
In this paper we concentrate on the weak coupling regime $\omega_k /\Omega_o 
\ll 1$ when the idea of an effective Hamiltonian is meaningful (we recall that
by definition $H_{eff}$ is operating in the low-energy subspace, and that a
consistent solution requires $\omega_k \ll \Omega_o$).
It is then fairly straightforward to handle interaction effects, in an
expansion  in $\omega_k / \Omega_o$. 

Since we want an effective Hamiltonian which describes both
$\vec{\tau}$ and the $\{ \vec{ \sigma }_k \}$, we examine, instead of the
transition amplitude $\Gamma^o_{\alpha \beta }$ in (\ref{z.1}), the 
more general  matrix element 
\begin{eqnarray}
 &&\Gamma_{\alpha \beta }( \{ \sigma_k^{(\alpha )}, \sigma_k^{(\beta )} \}; t) 
= \nonumber \\
\prod_{k=1}^N \int_{\vec{\sigma}_k^{(\alpha )}}^{\vec{\sigma}_k^{(\beta )}} 
{\cal D} \vec{\sigma}_k(\tau )& &
 \int_{\vec{n}_\alpha }^{\vec{n}_\beta } {\cal D} \vec{n}(\tau ) 
exp \left\{   -\int d\tau \big[ {\cal L}_o(\tau ) +
\sum_{k=1}^N{\cal L}_k^o(\tau ) +\delta {\cal L}_{\sigma}(\tau ) \big] 
\right\} \;.
\label{z.17}
\end{eqnarray}
\begin{equation}
{\cal L}_k^o(\tau ) = - {i \over 2} \dot{\theta }_k \varphi_k \sin \theta_k 
\label{z.18}
\end{equation}
\begin{equation}
\delta {\cal L}_{\sigma}(\tau ) = \sum_{k=1}^N \vec{\sigma}_k(\tau ) 
\cdot \vec{\gamma}_k(\tau ) \;.
\label{z.19}
\end{equation}
where $\vec{\gamma}_k(\tau )$ is now the time-dependent field 
\begin{equation}
\vec{\gamma}_k(\tau ) = {\omega_k \vec{S} ( \tau ) \over 2 S } \;,
\label{z.gamma}
\end{equation}
and ${\cal L}_o(\tau )$ was defined in (\ref{z.2}).
We assume in this calculation that the instanton is fast and the nuclear 
dynamics is slow and therefore we first solve for the instanton trajectory
minimising ${\cal L}_o(\tau )+ \delta {\cal L} (\tau )$ and calculate then
the transition amplitude from $\vert \vec{n}_\alpha  >$ to 
$\vert \vec{n}_\beta  >$. Since we deal here with non-diagonal matrix 
elements $(\alpha \ne \beta)$, we deal only with the actual transitions,
i.e., with the matrix elements $ \Gamma_{ \Downarrow \Uparrow }$.

The result of such a calculation will depend, of course, on the particular 
form of the giant spin Hamiltonian, the spin-bath Hamiltonian and the 
interaction potential. Let us therefore perform the calculation for the 
easy-axis/easy-plane model (\ref{1}) in zero external magnetic field 
coupled to the nuclear spins via hyperfine fields. The coupling term 
(\ref{z.19}) in the Lagrangian then takes the form
\begin{equation}
\delta {\cal L}_{\sigma} =  \sum_{k=1}^N 
{\omega_k \over 2} \bigg[ \sigma_k^x(\tau ) \sin \theta \cos \varphi +
\sigma_k^y(\tau ) \sin \theta \sin \varphi + 
\sigma_k^z(\tau ) \cos \theta \bigg] \;. 
\label{z.21}
\end{equation}
The analogy between this term and (\ref{h.99}) is obvious. Within the same 
accuracy, the correction to the minimal action from the hyperfine 
coupling is given by 
\begin{equation} 
\delta A_{\sigma}^{(\eta )} = \sum_{k=1}^N {\omega_k \over 2}\int d\tau 
\vec{\sigma}_k(\tau ) \cdot \vec{h}_{eff}^{(\eta )}(\tau ) =
\int d\tau \sum_{k=1}^N  \delta {\cal L}_k^{(\eta)} (\tau ) \;,
\label{z.24}
\end{equation}
where the instanton-generated field acting on the nuclear spins is defined as
\begin{equation} 
\vec{h}_{eff}^{(\eta )}(\tau ) = \left( \eta \sin \theta^{(\eta )} (\tau ) ,
\; i \eta \sqrt{K_{\parallel} / K_{\perp }}  \sin \theta^{(\eta )} (\tau )  ,
\; \cos \theta^{(\eta )} (\tau ) \! \right) \;,
\label{z.25}
\end{equation}
For the microscopic nuclear spin $1/2$ we do not use a quasiclassical 
method to evaluate the transition amplitude. Instead, we notice that the 
problem of evaluating the dynamics of $\sum_{k} [{\cal L}_k^o(\tau )+
\delta {\cal L}_k^{(\eta)} (\tau ) ]$ is identical to solving the spin 
rotation of each of the nuclear spins in the time dependent magnetic field 
$\omega_k \vec{h}_{eff}(\tau )/2$. The integrals (\ref{z.24}) are easy to 
evaluate for the instanton trajectory in the easy-axis/easy-plane model 
and we find [compare with (\ref{h.95})]
\begin{equation} 
\delta A_{\sigma}^{(\eta )} = \eta \sum_{k=1}^N {\pi \omega_k \over 2 \Omega_o}
(\hat{\vec{x}} + i \sqrt{K_{\parallel} / K_{\perp }} \hat{\vec{y}} )  
\cdot \vec{\sigma}_k(\tau ) \;.
\label{z.26}
\end{equation}

A note is in order here. We do not use the Wick rotation from imaginary 
to real time to calculate the evolution of the microscopic spin 
wave-function in the time-dependent instanton field $\vec{\gamma}_k(\tau )$.
To justify this we consider the text book  solution for a related problem. It is 
formulated as follows. A spin-1/2 particle passes through a 1-D barrier 
along the $\hat{\vec{y}}$ - direction. An external magnetic field $H_x$ is nonzero 
only inside the barrier. What is the spin structure of the transmission
coefficient  (say in the basis set of eigenfunctions of 
${\hat{\sigma}}_z$)? It is easily shown that one has 
$T=T_o \exp \{ -\alpha {\hat{\sigma}}_x \}$, where $T_o$ is the
transmission coefficient when the external field is zero, and
$\alpha$ (which depends on $H_x$) is {\it real}. For this reason, we conclude that 
the co-flip amplitudes have to be calculated in imaginary time,
which means that the evolution of the environmental spin does 
depend on whether the external field acting on it is classical or
originates from the tunneling motion of a combined system.

To get the final form of the non-diagonal terms in the effective
Hamiltonian we sum up the contributions 
from the external magnetic field and from the hyperfine coupling to
 the spin bath and  write the result for the transition amplitude as 
\begin{equation} 
\hat{ \Gamma}_{ \Downarrow \Uparrow }(t) = 
it \sum_{\eta =\pm } \sqrt{{2 \over \pi } Re A_o } 
\Omega_o \exp \left\{ -A_o^{(\eta )} - i
\eta \sum_k \alpha_k \vec{n} \cdot \hat{\vec{\sigma}}_k + i
\eta \beta_o \vec{n} \cdot \vec{H}_o \right\}  \;;
\;\;\; (\Omega_o^{-1} \ll t \ll \Delta_o^{-1} ) \;;
\label{z.28}
\end{equation}
\begin{equation} 
\alpha_k \vec{n} = {\pi \omega_k \over 2 \Omega_o} \big( -i
\hat{\vec{x}},~ 
 \sqrt{K_{\parallel} / K_{\perp }}~ \hat{\vec{y}} \big)  \;;
\label{z.29}
\end{equation}
\begin{equation} 
\beta_o \vec{n} = {\pi \gamma_e S \over \Omega_o} \big( -i
\hat{\vec{x}},~ 
 \sqrt{K_{\parallel} / K_{\perp }} ~\hat{\vec{y}} \big)  \;,
\label{z.291}
\end{equation}
where $\hat{\Gamma}_{ \Downarrow \Uparrow }$ is an operator in the 
nuclear spin subspace. We may now give the result for the non-diagonal 
coupling term in this easy-axis/easy-plane model 
\begin{eqnarray}
H_{eff}^{ND} &= &  {i \over t} \bigg[ \hat{\tau }_-  
\hat{\Gamma}_{ \Downarrow \Uparrow }(t)
     +  H.c. \bigg]  \nonumber \\
& = &  \bigg[  \hat{\tau }_- \Delta_o \sum_{\eta} \exp \left\{
- i \eta \left( \pi S + \sum_k \alpha_k \vec{n} \cdot \hat{\vec{\sigma }}_k
- \beta_o \vec{n} \cdot \vec{H}_o \right) \right\} + H.c. \bigg] \nonumber \\
& = & 2 \Delta_o \hat{\tau }_-  \cos \bigg[  
\pi S + \sum_k \alpha_k \vec{n} \cdot \hat{\vec{\sigma }}_k  -
 \beta_o \vec{n} \cdot \vec{H}_o  \bigg] + H.c. \;,
\label{z.30}
\end{eqnarray}
where $\Delta_o = -\sqrt{2 Re A_o/\pi  }~ \Omega_o ~ e^{-Re A_o}$ as before.
Adding the diagonal term (\ref{3.94}) we get the final truncated effective 
Hamiltonian:
\begin{eqnarray}
H_{eff} & = & \left\{ 2 \Delta_o \hat{\tau }_ -  \cos \bigg[ \pi S 
+ \sum_k \alpha_k \vec{n} \cdot \hat{\vec{\sigma }}_k  - 
 \beta_o \vec{n} \cdot \vec{H}_o  \bigg] + H.c. \right\} \nonumber \\
& + & {{\hat \tau }_z \over 2} \sum_{k=1}^N \omega_k^{\parallel} \:
 {\vec l}_k \cdot {\hat {\vec \sigma }}_k  + {1 \over 2} \sum_{k=1}^N 
\omega_k^{\perp}\: {\vec m}_k \cdot {\hat {\vec \sigma }}_k  \;.
\label{z.3301}
\end{eqnarray}
with the parameters $\vec{l}_k =\hat{\vec{z}}$, $\omega_k^\parallel
= \omega_k$, $\omega_k^\perp =0$, for the easy axis/easy plain
Hamiltonian (\ref{1}); the parameters $\alpha_k$, $\vec{n}_k$, and 
 $\beta_o$ are given in (\ref{z.29}) and (\ref{z.291}).

This result for the easy-axis/easy-plane model is not the most general
result we can get for the problem of a giant spin coupled to a nuclear spin
bath. More generally we will find that both the tunneling splitting 
$\Delta_o$ and the phase $\pi S$ will be renormalised by the coupling to the
nuclear spins - these renormalisations are $O(\omega_k^2 /\Omega_o^2 )$
for each nuclear spin (they correspond to a change in the instanton
trajectory caused by the coupling to the $\{ \vec{\sigma}_k \}$, and
hence appear first in the  2-nd order in this coupling). Thus more generally
we find that $\pi S \to \Phi $, and $\Delta_o \to \tilde{\Delta}_o$, where
\cite{our93,our-book}:
\begin{equation}
\Phi = \pi S + \sum_k \phi_k  \;,
\label{z.31}
\end{equation}
\begin{equation}
 \tilde{\Delta}_o = \Delta_o \exp \{ -\sum_k \delta_k \}  \;.
\label{z.32}
\end{equation}
Physically, the $\{ \phi_k \}$ are the Berry phase terms to be added to the
instanton phase, coming from the nuclei; and the 
$\{ \delta_k \}$ describe the  renormalisation of the tunneling
splitting. We do not attempt to calculate either $\phi_k$
or $ \delta_k $ here, since their effects are not going to be observable - 
we only observe $\tilde{\Delta}_o $ and $\Phi$, not $\Delta_o$ or the "bare"
phase $\pi S$.

We have now achieved our main aim in this section, of reducing the giant spin
Hamiltonian to a truncated one. It is interesting to notice what kinds
of process are now included in this effective Hamiltonian. Notice that
if we expand out the cosines in (\ref{z.301}), we see that we have a 
whole series of terms like
\begin{equation} 
\hat{\tau}_{\pm}  \Gamma_{\alpha \beta \gamma \delta \cdots }
\hat{\sigma}_{k_1}^{\alpha} \hat{\sigma}_{k_2}^{\beta}
\hat{\sigma}_{k_3}^{\gamma} \hat{\sigma}_{k_4}^{\delta} \cdots  \;.
\label{3.111}
\end{equation}
in which the instanton flip of the giant spin couples simultaneously
to many {\it different} nuclear spins. Thus, as well as the simple diagonal
and non-diagonal transitions, we also have the possibility (see Fig. 2) 
of multiple transitions in the bath, stimulated by a single instanton. 
The parameter $\lambda = {1 \over 2} \sum_k \vert \alpha_k \vert ^2$ 
- as already noted,  measures the average number of nuclear spins to
be flipped during each instanton. 


\section{Exact diagonalization studies.}
\label{sec:4}

In this section we wish to verify our analytical derivation (\ref{z.301})
of $H_{eff}$ by an  exact diagonalization method, and 
to analyse the  structure of the low-energy effective Hamiltonian
for arbitrary values of  parameters $S$ and $K_{\perp} / K_{\parallel}$, 
including regimes where the WKB approach breaks down.

For simplicity we consider  the  Hamiltonian for a giant spin
coupled to a single nuclear spin 1/2 (for small $\omega_k
/\Omega_o$ nuclear spin contributions are {\it additive}); a
similar tactic was employed by Abarenkova and Angles d'Auriac
\cite{d'Auriac}. We have  
\begin{equation} 
H(\vec{S} , \vec{\sigma}) = {1 \over S} \bigg[ -S_z^2 + \Lambda S_y^2 + 
{ \omega_o \over 2} \vec{S} \cdot \vec{\sigma}  \bigg] - 
\vec{H}_o \vec{S} \;,
\label{ed.1}
\end{equation}
where we measure energy in units of $K_{\parallel}$, magnetic field in
units $K_{\parallel} / \gamma_e$, and we define 
the anisotropy parameter $\Lambda =
K_{\perp}/K_{\parallel}$. Let us also apply the magnetic field 
$\vec{H}_o$ along the $\hat{\vec{y}}$ axis (i.e., $\vec{H}_o = \vec{H}_y$). 
For this particular choice of the external parameters we can rewrite 
(\ref{z.301}) as follows 
\begin{equation}
H_{eff} = \left\{ 2 \Delta_o \hat{\tau }_ -  \cos \bigg[ \pi S 
- \psi + \alpha \vec{n} \cdot \hat{\vec{\sigma }} \bigg] 
+ H.c. \right\} + 
{\omega_o \over 2} {\hat \tau }_z \cdot {\hat {\sigma }}_z +
{ \omega_o H_y \over 4 \Lambda } {\hat {\sigma }}_y
 \;,
\label{ed.2}
\end{equation}
where the last two terms describe the static 
diagonal interaction (and as before, we neglect both
the interaction between the nuclear spin and  magnetic field, and
the internuclear dipolar coupling). In this equation we have defined
\begin{equation}
\alpha \vec{n} = {\pi \omega_o \over 2 \Omega_o}
\big( -i \hat{\vec{x}}, ~
 \hat{\vec{y}}/\sqrt{\Lambda} \big)  \;, 
\label{ed.3}
\end{equation}
\begin{equation}
\psi = { \pi S H_y \over 2 \Lambda } \;.
\label{ed.fi}
\end{equation}

Let us start by defining  the diagonalization procedure. 
Since the giant spin only weakly couples to the nuclear spin 1/2, 
we have four lowest energy levels, well-separated from  higher
levels by the instanton frequency $\Omega_o$. The Hamiltonian
(\ref{ed.2}) is given in the Hilbert space of these four levels. 
Therefore, we can write
\begin{equation}
\langle a \vert H_{eff} \vert b \rangle \approx
\langle a \vert H \vert b \rangle \;;
\;\;\;\; (a \;, b = 1 \;, ... \;, 4) \;
\label{ed.4}
\end{equation}
or
\begin{equation}
\langle a \vert H_{eff} \vert b \rangle \approx 
\langle a \vert \sum_{\gamma=1}^4 E_{\gamma} \vert \Psi_{\gamma} \rangle
\langle \Psi_{\gamma} \vert b \rangle 
 =  \sum_{\gamma=1}^4 ( {\hat {U}^+} )_{a \gamma} E_{\gamma} 
{\hat{U}}_{\gamma b} \;,
\label{ed.5}
\end{equation}
where 
\begin{equation}
{\hat{U}}_{\gamma b} = \langle \Psi_{\gamma} \vert b \rangle \;
\label{ed.6}
\end{equation}
and the $\vert \Psi_{\gamma} \rangle $ are the four eigenfunctions of the 
Hamiltonian (\ref{ed.1}), corresponding to the four lowest energy levels
$ E_1 \;, ... \;, E_4 $. In fact, Eq.~(\ref{ed.4}) is nothing but
the definition of the effective Hamiltonian.

The choice of the 
appropriate basis set of $\vert a \rangle$ is important here.
The most convenient and physically transparent basis set is that
of the noninteracting system, i.e. the eigenfunctions of
$H=H_0(\vec{S}) $. 
Let $\vert \Phi_1^o \rangle$ and $\vert \Phi_2^o \rangle$ be the 
eigenfunctions corresponding to the two lowest levels of $H_o$ for
the giant spin $\vec{S}$. Then the functions 
\begin{equation}
\vert \chi_{\pm} \rangle = { \vert \Phi_1^o \rangle \pm \vert 
\Phi_2^o \rangle \over \sqrt {2} }
\label{ed.8}
\end{equation}
are localised single-well states, which in the semiclassical
limit correspond to the previously introduced 
$\vert \vec{n}_1 \rangle$ and $\vert \vec{n}_2 \rangle$. With
nuclear spins added we define the basis set $\vert a=1,2,3,4 \rangle $ as
\begin{equation}
\vert 1 \rangle = \vert \chi_+ \rangle \otimes {1 \choose 0}
\;, \;\;
\vert 2 \rangle = \vert \chi_+ \rangle \otimes {0 \choose 1}
\;, \;\;
\vert 3 \rangle = \vert \chi_- \rangle \otimes {1 \choose 0}
\;, \;\;
\vert 4 \rangle = \vert \chi_- \rangle \otimes {0 \choose 1} 
\label{ed.9}
\end{equation}
(where $\otimes$ stands for the direct product of state vectors).
This particular choice of the basis set $\vert \alpha \rangle$
is the starting point for our diagonalization procedure.

The diagonalization procedure itself is arranged as follows. First,
one  calculates the eigenstates  of the noninteracting problem to
construct the set $\vert a \rangle$, and then diagonalizes the
interacting Hamiltonian to obtain the energy spectrum and the
unitary transformation $\hat{U}$
as defined in Eqs.~(\ref{ed.5}) and  (\ref{ed.6}).
As $H_{eff}$ is a matrix $(4 \times 4)$ matrix it can be 
written  as
\begin{equation}
H_{eff} = \sum_{i ,j = 0}^3 C_{ij} \cdot 
{\hat{\tau}}_{i} \otimes {\hat{\sigma}}_{j} \;,
\label{ed.12}
\end{equation}
where ${\hat{\tau}}_{i}$ and $2 \cdot {\hat{\sigma}}_{i}$ 
($i = 1,2,3$) are the Pauli matrices and ${\hat{\tau}}_o$,  
${\hat{\sigma}}_o$ are the unit matrixes. Thus
\begin{equation}
C_{ij} = Tr \{ H_{eff} \cdot 
(\hat{\tau}_{i} \otimes \hat{\sigma}_{j}) \} \;.
\label{ed.13}
\end{equation}
We analyse these coefficients and compare them
with those resulting from (\ref{ed.2}). If $S$ is an integer we get
(we omit here the $\otimes $ sign)
\begin{eqnarray}
H_{eff} & \approx &
(-1)^S \left \{ 2 \Delta_o \cos{\psi} \cdot {\hat{\tau}}_x {\hat{\sigma}}_o 
- 2 \Delta_o \alpha  \sin{\psi} \cdot 
{\hat{\tau}}_y {\hat{\sigma}}_x 
+ 2 \Delta_o { \alpha \over \sqrt{\Lambda}}
 \sin{\psi} \cdot {\hat{\tau}}_x {\hat{\sigma}}_y
\right \}  \nonumber \\
& & + {\omega_o \over 2} {\hat \tau }_z {\hat {\sigma }}_z 
+ { \omega_o H_y \over 4 \Lambda } {\hat \tau }_o {\hat {\sigma }}_y \;;
 \;\;\;\;\;\;\; \alpha = {\pi \omega_o \over 2 \Omega_o } \;,
\label{ed.14}
\end{eqnarray}
and for $S=1/2$-integer we get
\begin{eqnarray}
H_{eff} & \approx & 
(-1)^{2S+1} \left \{ 
2 \Delta_o \sin{\psi} \cdot {\hat{\tau}}_x {\hat{\sigma}}_o + 
2 \Delta_o \alpha  \cos{\psi} \cdot 
{\hat{\tau}}_y {\hat{\sigma}}_x - 
2 \Delta_o { \alpha \over \sqrt{\Lambda}}
\cos{\psi} \cdot {\hat{\tau}}_x {\hat{\sigma}}_y
\right \} \nonumber \\
& & + {\omega_o \over 2} {\hat{\tau}}_z {\hat {\sigma }}_z +
{ \omega_o H_y \over 4 \Lambda } {\hat {\tau }}_o {\hat {\sigma }}_y \;.
\label{ed.15}
\end{eqnarray}

Now we turn to the results. 

\vspace{0.3cm}
{\bf (a) Semiclassical regime S $\gg$ 1}: We start by carrying out the 
numerical analysis for a 
wide region of parameters $S$ and $\Lambda$ and as an example we present
our results for $S=50$, $\Lambda=50$, and $\omega_o=0.2$ 
(recall again that energy is measured in units of $K_{\parallel}$).
In full agreement 
with (\ref{ed.14}) the effective Hamiltonian $H_{eff}^{ED}$, derived by 
the exact diagonalization, has only five non-zero coefficients  
$C_{ij}$, i.e.,
\begin{equation}
H_{eff}^{ED} = C_{xo} \cdot {\hat {\tau }}_x {\hat {\sigma }}_o + 
C_{xy} \cdot {\hat {\tau }}_x {\hat {\sigma }}_y + 
C_{yx} \cdot {\hat {\tau }}_y {\hat {\sigma }}_x + 
C_{oy} \cdot {\hat {\tau }}_o {\hat {\sigma }}_y + 
C_{zz} \cdot {\hat {\tau }}_z {\hat {\sigma }}_z \;.
\label{ed.16}
\end{equation}
In Fig. 4 we present the behavior of the 4 non-diagonal coefficients 
 as functions of the magnetic field $H_y$, and compare 
with the corresponding analytical forms (\ref{ed.14}),
coming from the instanton analysis. 
As to the value of $C_{zz}$, numerically it is 0.0972 in comparison with 
the analytical result 0.1, showing a small  zero point vibration
(z.p.v) reduction of 
$\langle \chi_{+} \vert \hat{S}_z \vert \chi_{+} \rangle <S$. 

We also have performed calculations for  half-integer spins. For 
$S=49.5$, $\Lambda=49.5$, and $\omega_o=0.2$, for example, we have 
found a similarly
good agreement between the numerical and analytical 
(\ref{ed.15}) results.

Bearing in mind that we kept only the leading terms in the 
hyperfine coupling when deriving
(\ref{ed.14}, \ref{ed.15}) we conclude 
from these results that the effective Hamiltonian 
(\ref{ed.2}) (and (\ref{z.301})) gives a complete description 
of the low-energy  dynamics of the system in weak magnetic field 
for $S \gg 1$, and that moreover the values of the coefficients 
computed for the effective Hamiltonian by the instanton expansion are in 
rather good agreement with the exact results. 

\vspace{0.3cm}
 {\bf (b) Beyond WKB: Arbitrary values of S and $\Lambda$}: 
Although the above results strongly support the validity
of our effective Hamiltonian (\ref{z.301}) for $S
 \gg 1$, it is of considerable interest to explore its validity
 when we are far from this WKB limit. In fact it was shown in the
 original papers of van Hemmen et al., as well as Enz
 and Schilling \cite{Hemmen}, that WKB results for the {\it free}
giant spin are
 remarkably accurate even when $S$ is less than 10, at least
as far as the calculation of tunneling splittings are
concerned. Thus we shall now use
 our exact diagonalisation results to tell us how accurate is our
 $H_{eff}$ in (\ref{z.301}), when $S$ and $\Lambda$ are arbitrary,
 and the action $A_o$ is not much larger than unity.
 
 We begin by defining a function $\Psi (S,\Lambda,H_y)$, such that
 the actual tunneling splitting $\Delta_o (\Psi ) =
 \Delta_o \cos \Psi $ for the non-interacting problem (for
 $S$-integer). In the semiclassical limit $\Psi = \psi $; we wish
 to see
 deviations for small $S$ and large $\Lambda$. Notice that even for
 small values of $S$, the coefficient $C_{x0}$ in (\ref{ed.16})
 still oscillates with $H_y$ like $cos \Psi $ (and 
 $C_{xy}$ and $C_{yx}$ like $\pm \sin \Psi$).

In Fig. 5 we present our numerical results for
$\Psi /\psi $ as a function of $S$ and $\Lambda$, with 
$H_y^o$ equal to some value smaller than one 
period of oscillation of $\Delta_o (\Psi )$ [in fact, in most cases 
$\Psi \sim H_y$ rather precisely unless $S <3$]. 
We plot the ratio $\Psi / \psi$ as a 
function of S at $\Lambda=50$ and $\Lambda=10$ (Fig. 5.a) and as a 
function of $\Lambda$ at $S=50$ (Fig. 5.b) to see where the analytical 
prediction (\ref{ed.fi}) becomes invalid. As we can see
from Fig. (5.a), the difference between the exact value of $\Psi$ 
and the semiclassical result $\Psi =\psi $ 
occurs at small values of $S$, whereas 
for $S > 10$ this difference has almost disappeared. Moreover, the 
curve at $\Lambda=10$ is falling to 1 more rapidly 
than at $\Lambda=50$, 
because with increasing  effective action $A_o$ the WKB 
description becomes more accurate. Fig.(5.b) shows the violation of the
WKB regime with increase of $\Lambda$ at $S=50$ as  
the effective action $A_o$ goes down.  At $S=50$ the analytical formula 
(\ref{ed.fi}) for
$\psi$ begins to show appreciable deviations from
 the exact result once $\Lambda \geq 700$.

So far so good; we now turn to the analysis of the 
interaction Hamiltonian itself, by 
considering the dependence 
of the coefficient $\alpha$ on $S$ and $\Lambda$. First, we calculate 
$\alpha$ as a function of $S$ at a constant value of the
effective action $A_o$. One expects that for large $A_o$ 
there should be good agreement between the  numerical result and
the analytical  result  (\ref{ed.3}). In Fig. 6 we plot the coefficient 
$C_{yx}$, divided by $2 \Delta_o \sin \Psi$, together with 
$\alpha$, defined in (\ref{ed.3}), vs the giant spin quantum number $S$ at 
$\Lambda =S^2 / 50$. With this definition of $\Lambda$ the effective action 
$A_o$ is a constant $\approx 14$. Since $\Delta_o$ and $ \Psi
(S,\Lambda ,H_y^o)$
are already known numerically, we plot the results of the  exact 
diagonalization for $\alpha$ together with the  analytical one.
From Fig.6 we see that once $S \ge 10$, the result (\ref{ed.3}) for
the effective coupling $\alpha$ is extremely good.

To study the deviations from (\ref{ed.3}), we show in Fig.7 
 numerical results for 
$C_{yx} / (2 \Delta_o \sin \Psi )$ together with $\alpha$ 
from (\ref{ed.3}) vs the giant spin quantum number at $\Lambda=50$ 
and $\Lambda =10$ (Fig. 7(a)) and vs $\Lambda$ at $S=50$ and $S=10$ 
(Fig. 7(b)).
The solid lines in Fig. 7(a) 
(parallel to the $S$-axis) are the analytical results. 
We can see, that at small values of the giant spin (at $S \leq 10$ in this 
particular case) it becomes useful to derive numerical results to describe its 
interaction with the nuclear spin. Notice (Fig. 7(b)) that the numerical 
(triangles) and analytical (solid) curves coincide extremely well for $S=50$ 
whereas for $S=10$ the differences between the numerical and analytical 
curves become obvious once $\Lambda \ge 100$.

Thus we can conclude that (\ref{ed.3}) becomes more and more accurate with 
the increase of $S$ at any fixed value of $\Lambda$ once the 
effective action $A_o=2 S / \Lambda^{1/2}$ becomes significantly
greater than unity (compare, e.g., in 
Fig. 7(b), the point $\Lambda=500$ for $S=10$ (triangles down), where  
the effective action $A_o < 1$ , with the point having 
the same value of $\Lambda$, but $S=50$, for which  
$A_o > 1$.

Finally we analyse the effect of   z.p.v.
 of the giant spin
on the value of the diagonal coupling. 
The orientation of the giant spin is fluctuating
around the classical energy minima, resulting in a reduction of the static 
diagonal interaction. It is easy to see from (\ref{z.6}) that the value
of $ < \sin^2{\theta} > \approx 1/A_o$. From this it follows
that the coefficient $\omega_o^{\parallel}$ in the longitudinal diagonal 
interaction (ie., that proportional to 
${\hat \tau }_z \cdot {\hat {\sigma }}_z$ in 
(\ref{ed.2})), can be corrected by replacing $\omega_o \Rightarrow 
\omega_o  (1 - 1/A_o)^{1/2}$. 
Thus, with increasing $\Lambda$, the giant spin 
becomes more and more "delocalized", and the extent of the
delocalisation is parametrized in the exact diagonalisation
results by the z.p.v. reduction of $C_{zz}$. To show how
accurate is the simple replacement $\omega_o \to \omega_o (1 -
1/A_o)^{1/2}$,
we compare in Fig.8 the coefficient $2 C_{zz} /\omega_o $ against
$(1 -1/A_o)^{1/2}$, both plotted as functions of $S$ 
at $\Lambda=50$ (triangles up) and at
$\Lambda=10$ 
(triangles down). 

\vspace{0.3cm}
 {\bf (c) Zero magnetic field: Nonlinear corrections in
$\alpha$}:~~ Our instanton derivation 
in section III was carried out only as far
as 1st-order in the effective coupling $\alpha_k$.
However we see immediately that in (\ref{ed.2}), when $S=$integer
and in zero external field, the nondiagonal term is affected by the
coupling only in the second order in $\alpha $, i.e., 
$\Delta_o \cosh  (\alpha ) \approx \Delta_o (1+\alpha^2/2)$.
As discussed above [see Eq.~(\ref{z.32})] in a consistent
calculation one may also have corrections $\sim \alpha^2 $ which
are not accounted for in Eq.~(\ref{ed.2}). To find these
corrections
we have performed the necessary numerical calculations for 
for integer and half-integer $S$ in zero external
field.

In Fig.9(a) we show the behavior of the coefficient $C_{xo} / 2 \Delta_o$ 
(or tunneling splitting renormalisation) 
vs $\alpha$ for integer $S$. Our analysis shows that the 
behavior of $C_{xo}$ in the region $0 \le \alpha \le 0.5$ is very close to 
$2 \Delta_o (1-\alpha^2/2)$. These results suggest that the
parameter $\delta$ in Eq.~(\ref{z.32}) is given by $\alpha^2$, and
the nonlinear correction may be accounted for by substituting
$\Delta_o \Rightarrow \Delta_o(\alpha ) = \Delta_o (1-\alpha^2)$
in the effective Hamiltonian (\ref{ed.2}).
In Fig 9(a) we plot the numerical results for 
$C_{xo} / 2 \Delta_o$ (triangles up) together with the analytical 
approximation $(1-\alpha^2) \cosh (\alpha )$ (solid line). All other 
coefficients $C_{ij}$ (except $C_{zz}$) are zero.

To see whether the ansatz $\Delta_o \Rightarrow \Delta_o(\alpha )$
really works well, we considered also the case of half-integer $S$.
 In Fig.9(b) we plot the coefficient $C_{yx} / 2 \Delta_o$ vs 
$\alpha$ (triangles up) together with the analytical approximation 
$(1-\alpha^2) \sinh (\alpha )$ (solid line). As we can see from 
this figure, our approximation works quite well in the region $0 \le \alpha 
\le 0.5$. 

This completes our numerical analysis of the low-energy effective 
Hamiltonian in the problem of the giant spin coupled to the spin enviroment.
Our main conclusion in this section can be formulated as follows: the 
predictions about the low-energy dynamics of the system in the magnetic 
field, given by (\ref{ed.2}) (and (\ref{z.301}) correspondingly),
are rather accurate when the effective action $A_o > 1$.


\section{Conclusion.}
\label{sec:5}
To conclude, it may be useful to summarize some of the esential mathematical 
features of our analysis. We started with a "high-energy" form for the central
spin coupled to its spin bath environment, given by (\ref{a.1}). This was
truncated to the low-energy form (\ref{z.301}) by an instanton-type procedure.
However the results of this truncation were different from the usual quantum 
environment models, mainly because the spin bath modes are strongly coupled 
to the central spin. As a consequence much of the physics of the problem is 
contained in dynamic non-diagonal terms (in contrast to the oscillator bath 
models, where such terms are almost irrelevant), as well as in almost static 
but very strong diagonal couplings. The instanton derivation also allows one
to evaluate the coefficients in the low-energy effective Hamiltonian,
for a given initial "high-energy" Hamiltonian for a giant spin coupled to 
a nuclear spin bath.

In the present paper we set out to find out how accurate this instanton
result for the low-energy effective Hamiltonian really is. Using an exact 
diagonalisation procedure, we were not only able to verify that the
effective Hamiltonian is correct, but also that the coefficients in it
are rather accurately given by the "first-order" (in $\omega_k /\Omega_o$)  
instanton expansion, provided we are working in the regime where the 
semiclassical WKB approximation works. 

The importance of this result is simply that the low-energy effective 
Hamiltonian, in the form we have given, turns out to be remarkably  
easy to use in practical calculations of the spin dynamics of the central
spin, as we have shown in a number of other papers (see, eg., ref.
\cite{JLTP}, and refs therein). From the present calculations we now have
a good idea of how trustworthy these calculations are- essentially they are
as good as the semiclassical approximation itself.


\section{Acknowledgement}
This work was supported by NSERC in Canada and by the Russian Foundation for
Basic Research (grant 97-02-16548). We would like to thank 
B.G. Turrell for some very helpful discussions, and also 
V. Kashurnikov for his help in writting of the exact diagonalization 
code. We also thank the Institute for Nuclear Theory at the
University of Washington for its hospitality and the Department of
Energy  for partial support during the completion of this work.

\begin{center}
{\bf FIGURE CAPTIONS }
\end{center}

\figure{{\bf Figure 1 }A typical trajectory for $S_z$, contributing 
to the path 
integral  for the propagator of a free giant spin. The time scale 
for "fast" variations of 
$S_z$ while it is in one of the 2 wells or making a transition
between the weells is $\Omega_o^{-1}$,
whereas the time between jumps (instantons) is much longer, roughly 
$\Delta_o^{-1}$.}

\figure{{\bf Figure 2 }Processes which may occur 
when a 2-state system is  coupled to some background
environment - we show some possible kinds of coupling between the path of the
2-state system (solid line) and the environment. There is a diagonal 
coupling (ie., a coupling
to $\tau_z$), indicated by "D", to a single environmental mode (whose 
propagator is indicated by a wavy line). The first non-diagonal coupling 
(labelled by "ND") is to a {\it single} environmental mode, whereas the 
second one couples simultaneously to 3 different environmental modes.

\figure{{\bf Figure 3 }Processes occuring in the propagator for a 2-state 
system coupled  to a set of oscillator modes, after one averages over these 
modes ("integrates them out"), which allows us to join the wavy lines of the
previous figure in the usual way.
Process (a) involves the excitation of an environmental mode when the
central spin is in state $|\Uparrow>$, and with re-absorption when it is in
state $|\Downarrow>$; process (b) has the central spin in the same state for
both emission and re-absorption. Both these processes involve only diagonal 
operators. Process (c) involves non-diagonal terms, and process (d) involves 
both diagonal and non-diagonal terms. 

\figure{{\bf Figure 4 }The numerical coefficients $C_{ij}$ 
(\ref{ed.13}) of the effective Hamiltonian (\ref{ed.16}) in comparison 
with the analytical expressions from (\ref{ed.14}). The figures are: 
(a) $C_{x0}$ (solid line) and $2 \Delta_o \cos{\psi}$ (dashed line), 
(b) $C_{xy}$ (solid line) and $2 \Delta_o \alpha \sin{\psi}/
\Lambda^{1/2}$ (dashed line), 
(c) $C_{yx}$ (solid line) and $  2 \Delta_o \alpha \sin{\psi}$ (dashed line), 
(d) $C_{y0}$ (solid line) and $\omega_o H_y / 4 \Lambda$ (dashed line); 
$S=50, \Lambda=50, \omega_o=0.2$.

\figure{{\bf Figure 5 }The ratio $\Psi (S,\Lambda,H_y^o) / \psi$ of the 
exact phase $\Psi $ to the weak-field expression for $\psi$ (\ref{ed.fi}): 
(a) vs the giant spin quantum number $S$ at $\Lambda=50$ (triangles up) and 
at $\Lambda=10$ (triangles down), (b) vs $\Lambda$ at $S=50$ (triangles up); 
$\omega_o=0.1$, $H_y^o=\Lambda / 8 S$.

\figure{{\bf Figure 6 }The numerical coefficient $C_{yx} / (2 \Delta_o 
\sin \Psi)$ (triangles up) together with the analytical co-flip amplitude 
$\alpha$ from (\ref{ed.3}) (solid line) at $\Lambda=S^2 / 50$; $\omega_o=0.1$,
$H_y^o=\Lambda / 8 S$.

\figure{{\bf Figure 7 }The deviation of $\alpha$ from the analytical 
expression (\ref{ed.3}). The figures are: (a) The numerical coefficient 
$C_{yx} / (2 \Delta_o \sin \Psi )$ at $\Lambda=50$ (triangles up) and at 
$\Lambda=10$ (triangles down) together with the analytical $\alpha$ 
from (\ref{ed.3}) (solid lines) vs the giant spin quantum number $S$, 
(b) The numerical coefficient $C_{yx} / (2 \Delta_o \sin \Psi )$ at $S=50$ 
(triangles up) and at $S=10$ (triangles down) together with the analytical 
$\alpha$ (solid line) vs $\Lambda$; $\omega_o=0.1$,
$H_y^o=\Lambda / 8 S$.

\figure{{\bf Figure 8 }The numerical coefficient $C_{zz} / 0.5 \cdot 
\omega_o$, vs the giant spin $S$ at $\Lambda=50$ (triangles up) 
and at $\Lambda=10$ (triangles down); $\omega_o=0.1$,
$H_y^o=\Lambda / 8 S$.

\figure{{\bf Figure 9 }Zero magnetic field: (a) The numerical coefficient 
$C_{x0} / 2 \Delta_o$ vs $\alpha$ at $S=50$ and $\Lambda=50$ (triangles up)
together with the analytical approximation $(1-\alpha^2)  \cosh(\alpha)$
(solid line); (b) The numerical coefficient $C_{yx} / 2 \Delta_o$ vs $\alpha$ 
(triangles up) at $S=49.5$ and $\Lambda=49.5$ together with the analytical 
approximation $(1-\alpha^2)  \sinh(\alpha)$ (solid line).


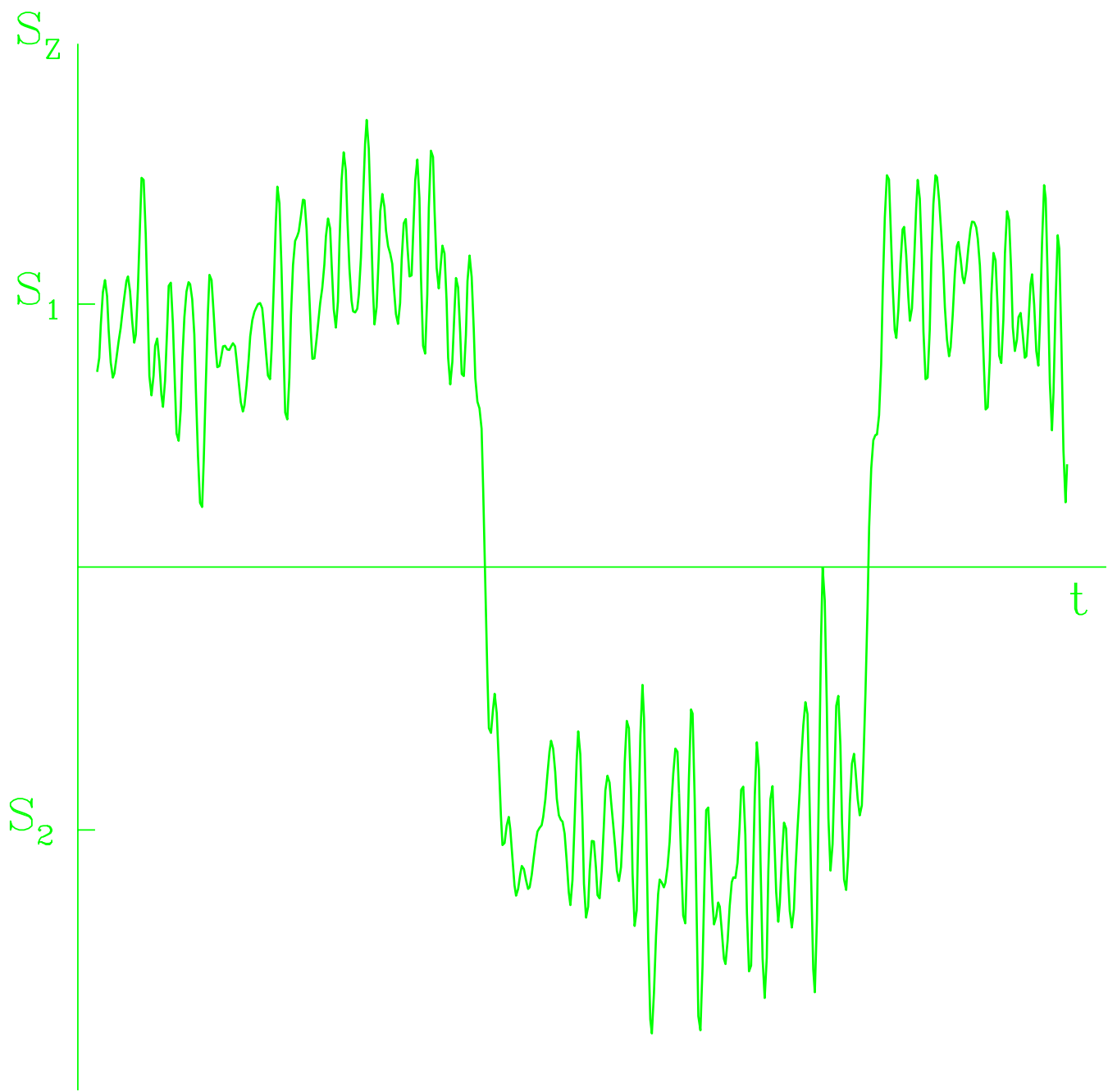
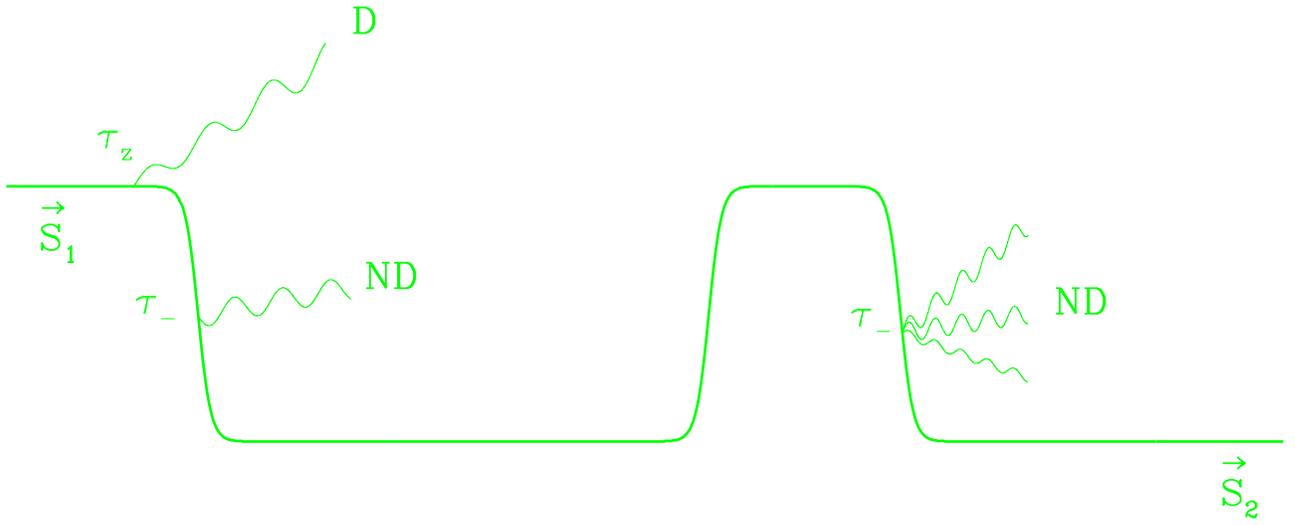
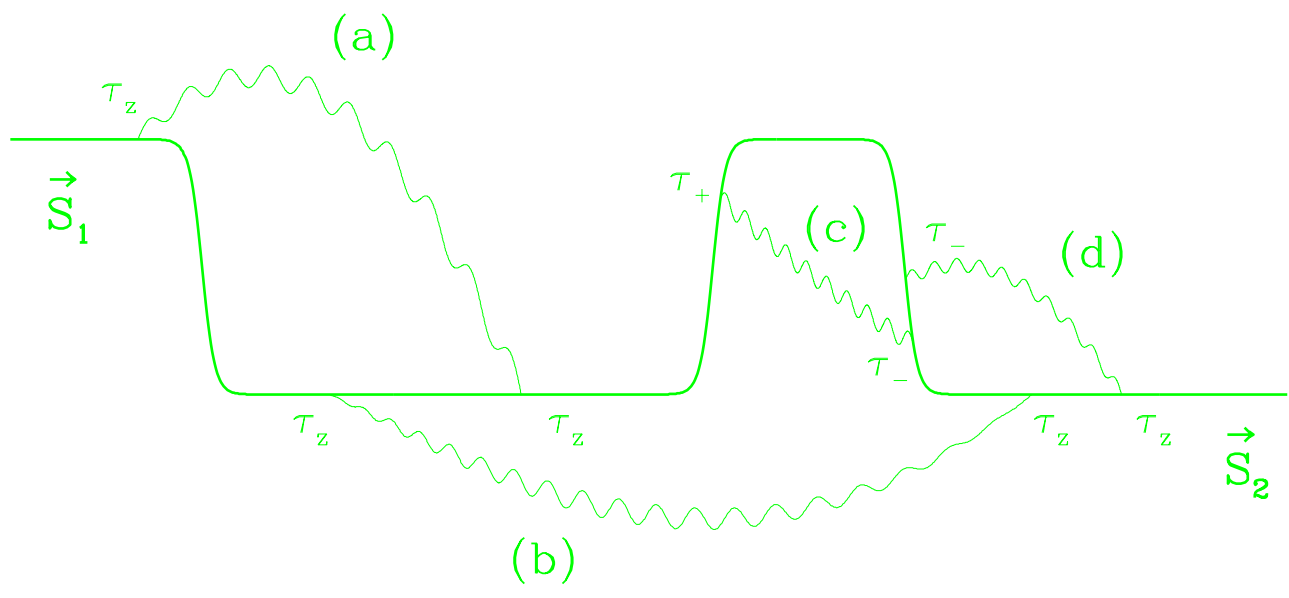
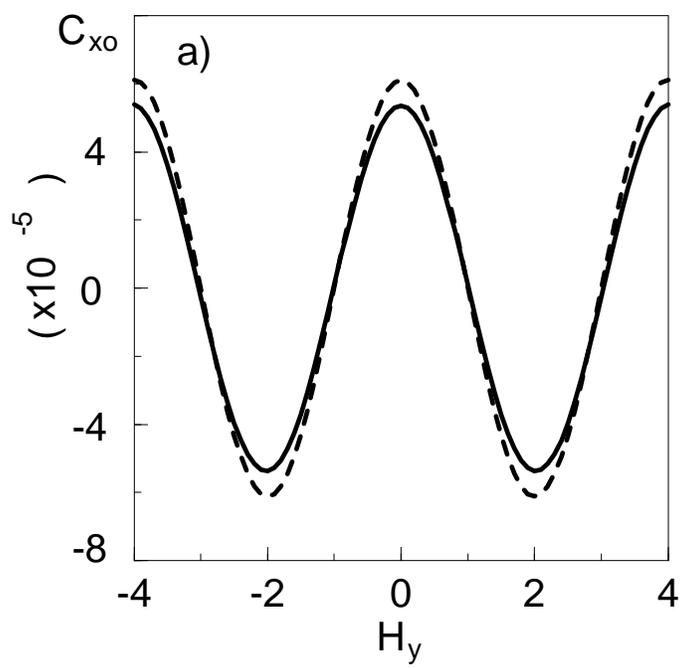
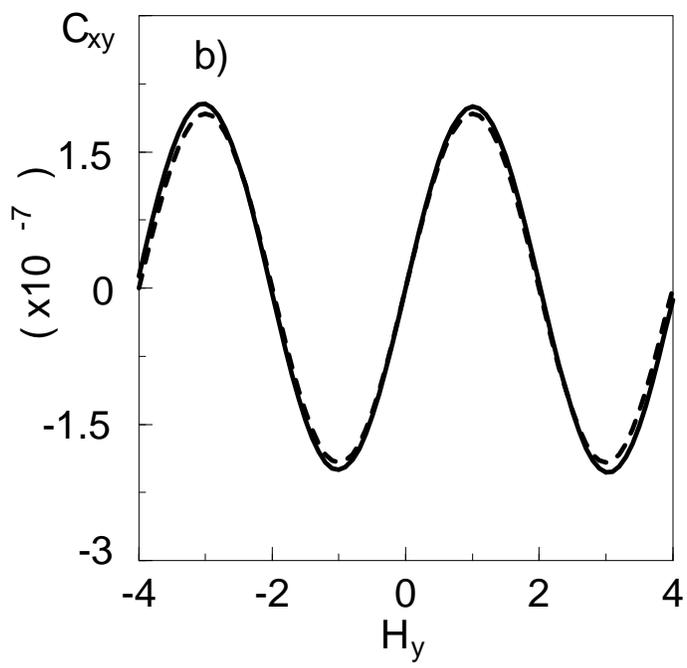
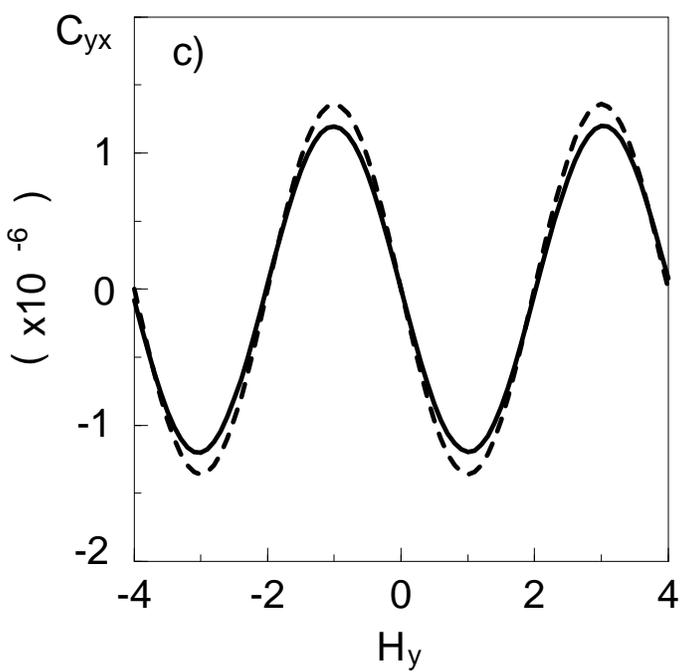
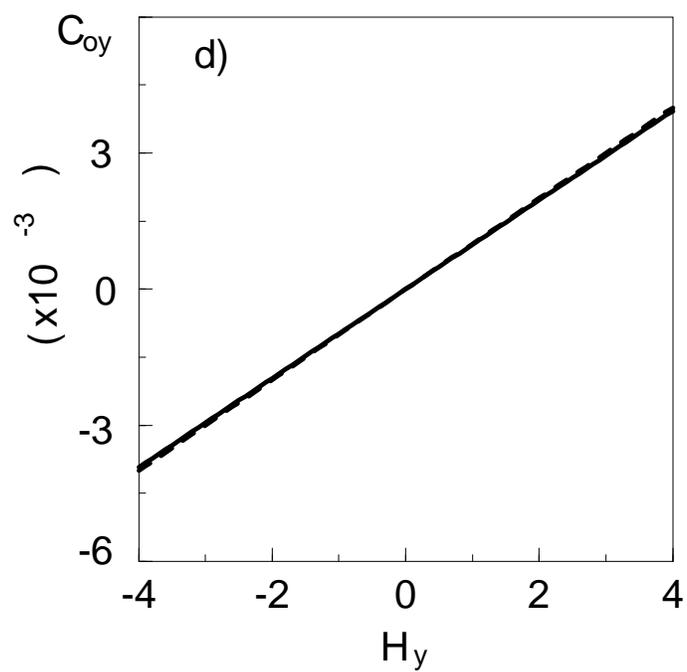
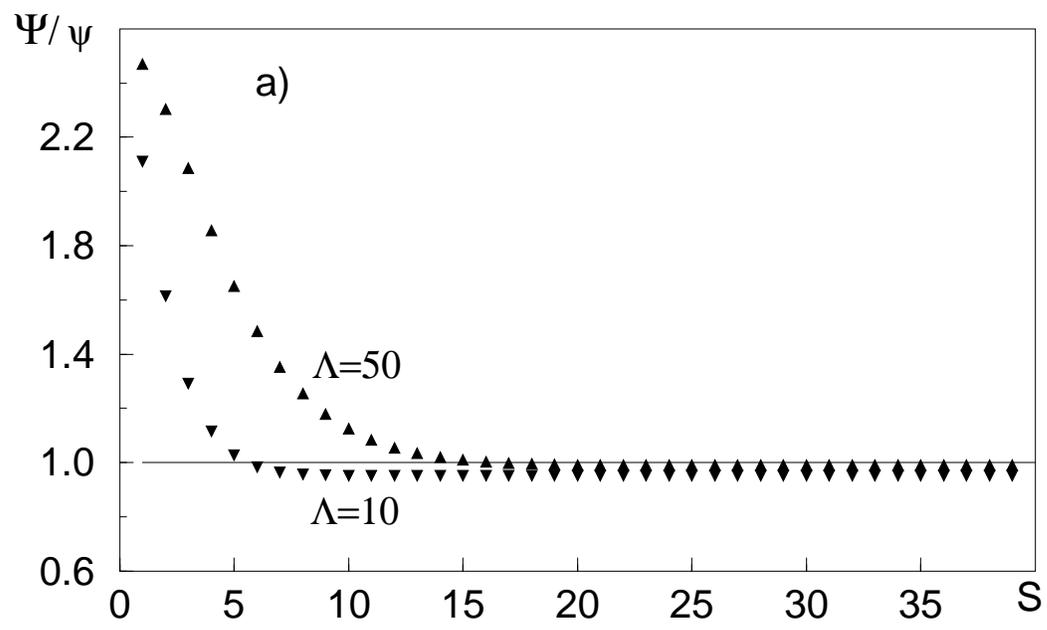
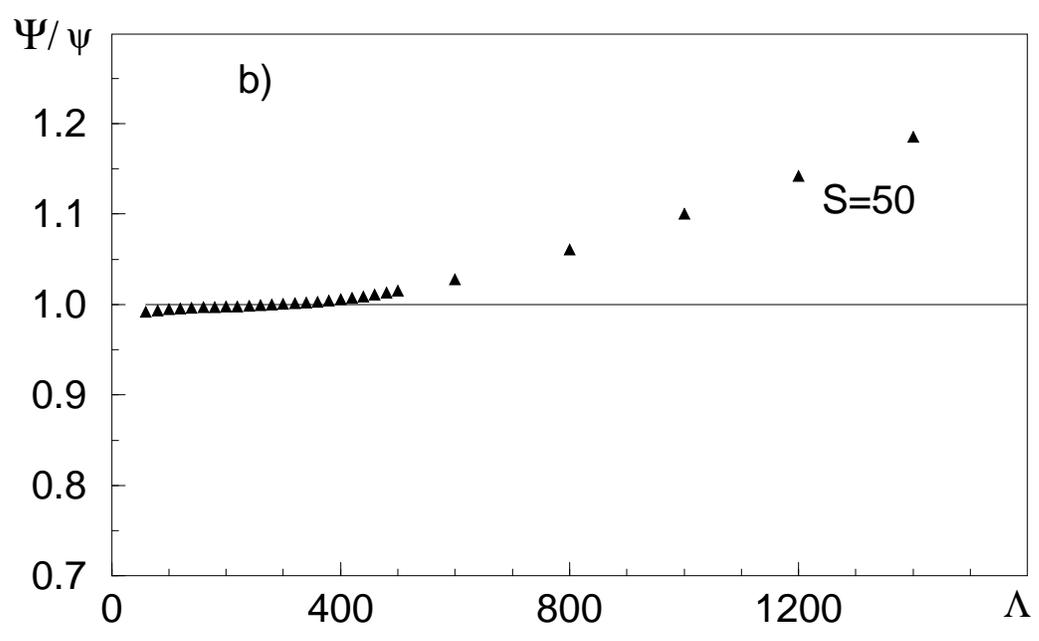
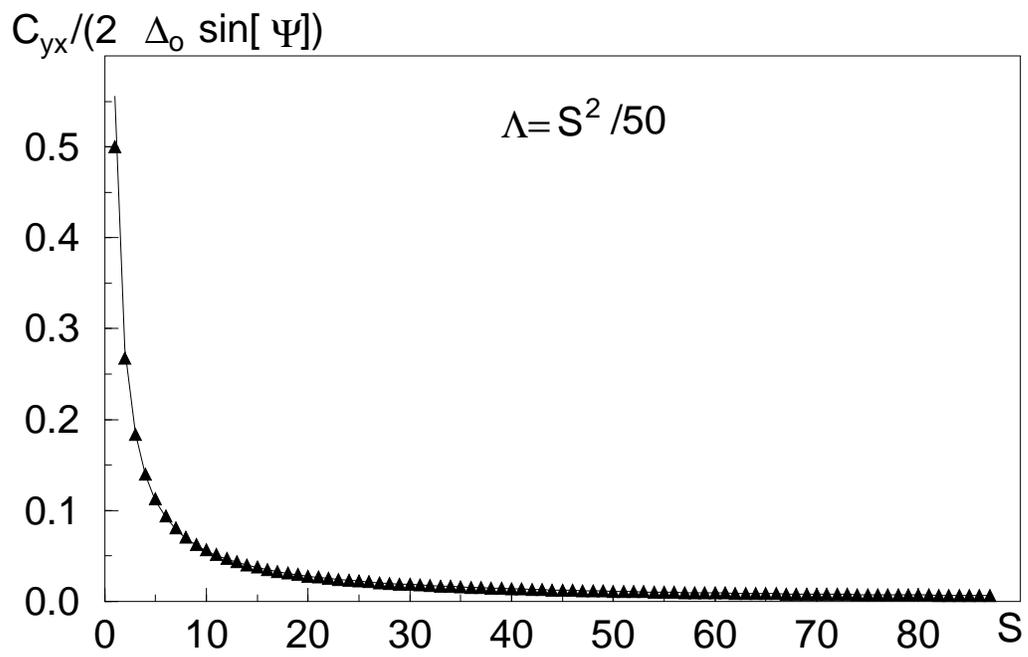
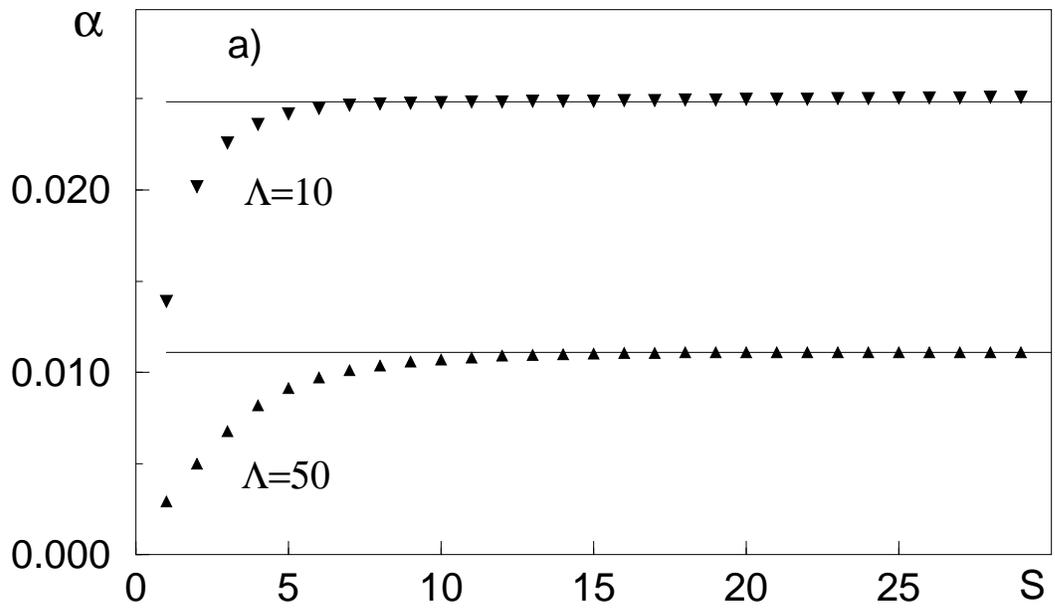
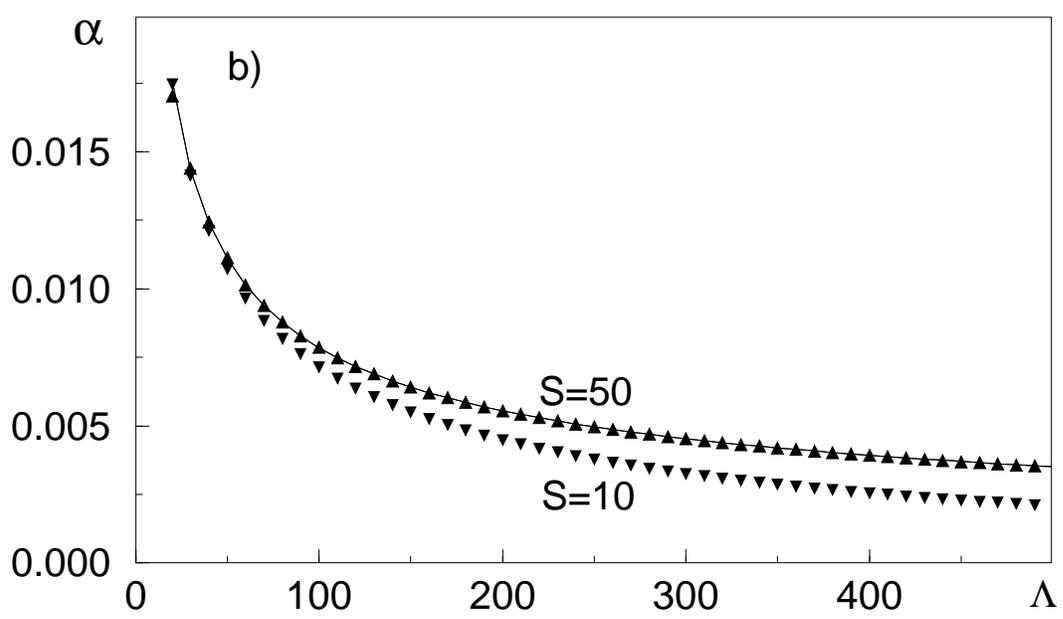
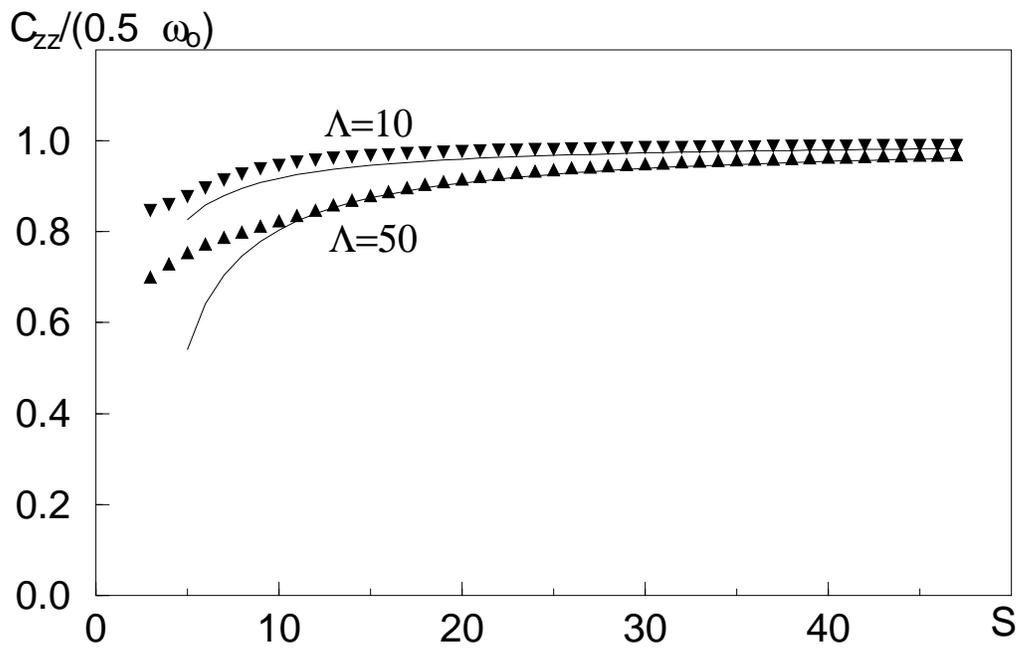
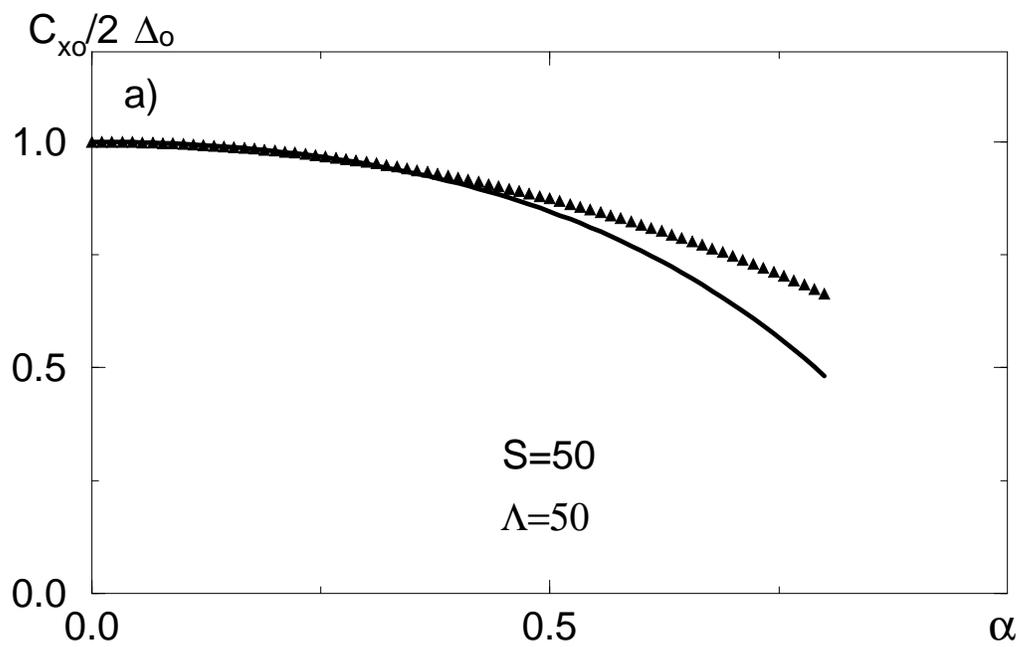
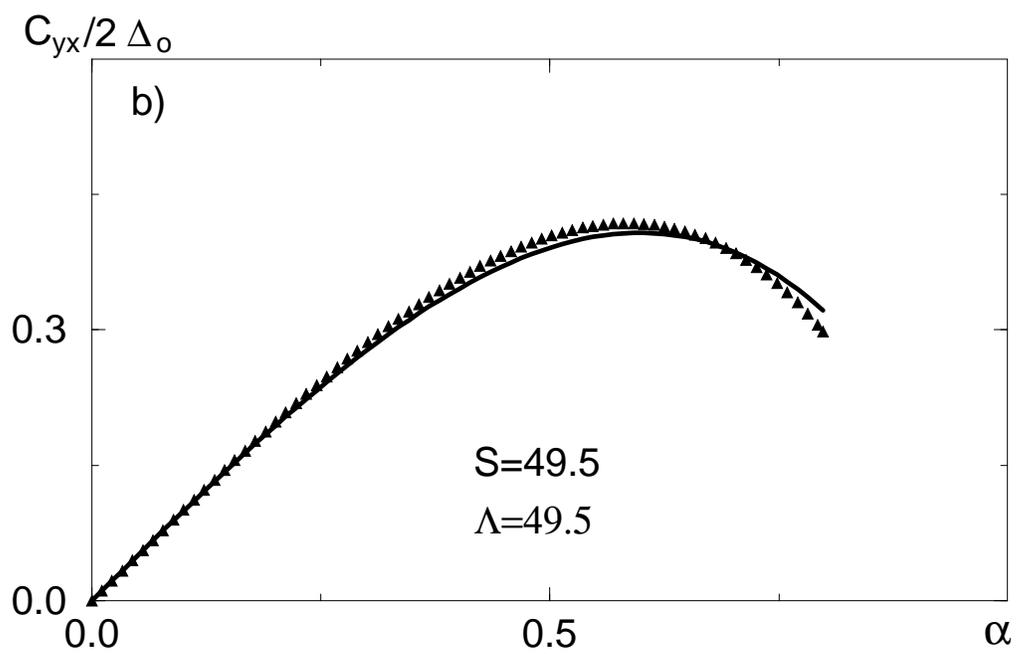
\end{document}